\documentclass[12pt,preprint]{aastex}
\begin{document}

\parindent=1.0cm

\title{
Intermediate Mass Early-type Disk Galaxies in the Virgo Cluster. II. Near-Infrared
Spectra and Evidence for Differences in Evolution \altaffilmark{1} \altaffilmark{2} \altaffilmark{3}}

\author{T. J. Davidge}

\affil{Dominion Astrophysical Observatory,
\\Herzberg Astronomy \& Astrophysics Research Center,
\\National Research Council of Canada, 5071 West Saanich Road,
\\Victoria, BC Canada V9E 2E7\\tim.davidge@nrc.ca; tdavidge1450@gmail.com}

\altaffiltext{1}{Based on observations obtained at the Gemini Observatory, which
is operated by the Association of Universities for Research in Astronomy, Inc., under 
a cooperative agreement with the NSF on behalf of the Gemini partnership: the National
Science Foundation (United States), the National Research Council (Canada), CONICYT
(Chile), Minist\'{e}rio da Ci\^{e}ncia, Tecnologia e Inova\c{c}\~{a}o (Brazil) and 
Ministerio de Ciencia, Tecnolog\'{i}a e Innovaci\'{o}n Productiva (Argentina).}

\altaffiltext{2}{This research has made use of the NASA/IPAC Infrared Science Archive,
which is operated by the Jet Propulsion Laboratory, California Institute of Technology,
under contract with the National Aeronautics and Space Administration.}

\altaffiltext{3}{This research makes use of the Virgo cluster galaxy database 
of McDonald et al. (2011).}

\begin{abstract}

	We discuss near-infrared (NIR) slit spectra of six 
early-type disk galaxies in the Virgo Cluster that were examined previously 
at visible/red wavelengths by Davidge (2018a, AJ, 156, 233). Despite having 
similar intrinsic luminosities, colors, and morphologies, the NIR 
spectrophotometric properties of these galaxies indicate that they are not 
a homogeneous ensemble differing only in terms of luminosity-weighted age 
and metallicity. While the depth of the CO(2,0) band is consistent with the 
centers of these galaxies having solar or slightly sub-solar luminosity-weighted 
metallicities, galaxy-to-galaxy variations in the depth 
of the NaI $2.21\mu$m doublet are found, with NaI$2.21\mu$m lines in three galaxies 
(NGC 4491, NGC 4584, and NGC 4620) that are deeper than those predicted for a 
solar chemical mixture and a solar-neighborhood mass function. In 
contrast, the CaI$2.26\mu$m triplet shows good galaxy-to-galaxy agreement, but 
tends to be deeper than the model prediction. Considering the depth of the NaD lines 
found by Davidge (2018a), the deep NaI$2.21\mu$m lines are tentatively attributed to a 
bottom-heavy mass function. This is counter to observed trends between mass function 
slope and velocity dispersion, and so the possibility of a super-solar [Na/Fe] 
is also discussed. Two of the three galaxies with deep NaI$2.21\mu$m (NGC 4584 and 
NGC 4620) have Sersic exponents that are consistent with a classical bulge. 
As for NGC 4491, its central NIR spectrum contains prominent 
emission lines. The relative strengths of Br$\gamma$ and H$_2$S(1), 
the presence of [FeII] emission, and the mid-infrared spectral-energy distribution are 
all consistent with intense recent star formation near the 
center of that galaxy. The NIR spectrum of NGC 4584 is devoid of line emission in the 
NIR, suggesting that star formation does not power the emission detected at visible 
wavelengths from that galaxy. Wavelengths that contain the Ballick-Ramsey C$_2$ band 
at $1.76\mu$m are matched by moderately metal-poor E-MILES model spectra with an age of 
2 Gyr. The radial age trends in these galaxies are in the opposite sense to those in 
late-type disk galaxies, and it is concluded that they 
have evolved in a cluster environment for at least several Gyr.

\end{abstract}


\section{INTRODUCTION}

	Environment plays a major role in defining the star formation histories 
(e.g. Butcher \& Oemler 1984; Hogg et al. 2004; Kauffmann et al. 2004) and morphologies 
(e.g. Dressler 1980; Postman \& Geller 1984; Moore et al. 1996; 
Dressler et al. 2007) of galaxies. As the nearest large 
galaxy cluster, Virgo is an unprecedented laboratory for exploring galaxy evolution in  
a dense environment. In addition to being nearby by cosmic standards, Virgo 
is also dynamically young (e.g. Tully \& Shaya 1984; Kim et al. 2016), and so studies of 
its constituent galaxies may provide information on the early and intermediate stages of 
evolution in a cluster environment. This in turn may yield insights into conditions 
that prevailed in distant clusters that formed during earlier epochs.

	A complicating factor for understanding galaxy evolution in Virgo is that there is 
sub-structure (e.g. Bingelli, Tammann, \& Sandage 1987; Bohringer et al. 1994), 
indicating that the cluster did not form as a monolithic entity. The presence of 
sub-structure is not surprising, as simulations suggest that large 
clusters likely form through the accretion of smaller structures (e.g. McGee et 
al. 2009). This has implications for the properties of present-day cluster galaxies as these 
may have originated in a diverse range of environments and evolved 
for many Gyr in settings that are very different from those in
the large cluster that they subsequently join (e.g. De Lucia 
et al. 2012; Wetzel et al. 2013; Taranu et al. 2014). Many basic galaxy properties -- such as 
the chemical mixture imprinted in the main mass of stars that form during early epochs and 
the orientation of kinemetic axes -- may already have been defined prior to entry into 
the larger cluster. Models examined by Hou et al. (2014) suggest that 
the effects of such `pre-processing' is likely significant for halo masses $> 10^{14.5}$ 
M$_{\odot}$. As the mass of the Virgo cluster is $8 \times 10^{14}$M$_{\odot}$ 
(Karachentsev et al. 2018), then pre-processing in sub-groups might have been significant 
among Virgo galaxies. Indeed, the kinematic properties of early-type galaxies 
in Virgo show evidence for accretion along two axes that likely trace 
distinct filaments along which the galaxies in Virgo formed (Kim et al. 2018). 

	Interactions with an ambient cluster medium are 
thought to deplete and disrupt the interstellar mediums (ISMs) 
of disk galaxies (e.g. Gunn \& Gott 1972; Rudnick et al. 2017), although 
gas reservoirs that feed the ISMs may be replenished in some cases (Grootes et al. 2017). 
It is thus not surprising that many disk galaxies in the Virgo cluster have 
properties that differ from their field counterparts. Disk galaxies in Virgo that are 
deficient in HI have ISMs that are more metal-rich than those in field galaxies, possibly 
due to strangulation of the gas supply (e.g. Skillman et al. 1996). Many gas-deficient 
Virgo spirals have truncated H$\alpha$ disks (e.g. Koopmann \& Kenney 1998; Koopmann 
et al. 2006; Fossali et al. 2013; Bodelli et al. 2020), often 
with undisturbed disks and `normal' star formation rates (SFRs) within the truncation 
radius, although there are exceptions (e.g. Koopmann \& Kenney 2004a). Star formation 
outside of the truncation radius appears to have stopped $\sim 0.5$ Gyr ago 
in some systems (Crowl \& Kenney 2008). As this is less than the cluster crossing time for 
Virgo then the cessation of star formation in the outer disk of at least 
some galaxies may not be related directly to a passage through the cluster core (e.g. Crowl 
\& Kenney 2008). Still, the age profiles of disk galaxies in Virgo appear to be related 
to environment (Roediger et al. 2012), highlighting the role that 
the present-day cluster plays in defining the evolution of disk galaxies.

	Passively evolving red spiral galaxies and lenticular galaxies are possible 
outcomes of evolution in a dense environment (e.g. Bamford et al. 2009). 
In fact, low mass anemic spiral galaxies in the nearby Universe appear to be found 
preferentially in the Virgo cluster, highlighting the role that environment plays 
in forming those systems (Fraser-McKelvie et al. 2018).
Number counts suggest that the transition from star-forming to passive disk galaxy is 
a process that spans many Gyr (Schawinski et al. 2014).
While the stripping of the ISM and strangulation of the gas supply are possible 
mechanisms for forming anemic red spirals on such a timescale, it has 
also been suggested that galaxy-galaxy interactions may play the dominant role 
in altering the evolution of otherwise blue disk galaxies in clusters (e.g. Boselli 
\& Gavazzi 2006). There is also evidence that factors other than environment 
may be at work. For example, the high incidence of Seyfert/LINER activity and 
bars when compared with bluer disk galaxies suggest that secular processes may play 
a role in the evolution of anemic red spirals (Masters et al. 2010). 
Disk truncation may also bias morphogical classification, in the sense of producing 
galaxies that might be classified as morphological type Sa but that lack large bulges
(e.g. Koopmann \& Kenney 1998, 2004b).

	When considered as a group, lenticular galaxies show diverse star formation 
histories (SFHs), with properties that depend on galaxy luminosity, which is a proxy 
for mass. Whereas massive S0 galaxies tend to be predominantly old, 
their dwarf counterparts have a significant younger stellar content 
(Barway et al. 2013), suggesting differences in evolutionary paths.
While classic S0 galaxies are usually associated with high-density 
environments, and hence may be the result of cluster-based processes 
such as stripping, strangulation, and galaxy-galaxy interactions, low 
mass lenticular galaxies can be found in lower density environments in the nearby 
universe. NGC 5102 and NGC 404 are among the closest examples of low mass 
lenticular systems. Both are relatively isolated and contain central populations with 
ages $< 1$ Gyr (Seth et al. 2010; Davidge 2015a). The radial metallicity 
profile of NGC 5102 suggests that the current morphology might be the 
result of the buckling of the bar in an LMC-like galaxy (Davidge 2015a). NGC 55 is a nearby 
irregular galaxy in a low density environment that may be transitioning 
into a dwarf lenticular morphology (Davidge 2018b; 2019). If moderate mass 
early-type disk galaxies in the Virgo cluster and the field have been subject to similar 
evolutionary processes then the former might show evidence 
of (1) a young stellar component, and (2) prolonged evolution in 
an environment that has a lower density than the present-day Virgo cluster.

\subsection{A Sample of Early-Type Intermediate Mass Disk Galaxies in Virgo}

	Davidge (2018a) discusses visible and red spectra of six intermediate mass 
early-type disk galaxies in Virgo. Various observational properties of these galaxies are 
listed in Table 1. The $g-i$ colors in this table have not been corrected for internal 
or line-of-sight extinction. The absolute $K$ magnitudes in Table 1 are consistent with 
total stellar masses that are probably comparable to -- or slightly larger than -- that 
of M33, for which M$_K \sim -20.4$ (Jarrett et al. 2003) assuming a distance modulus 
of 24.5 (Lee et al. 2002).

	The properties of the galaxies in Table 1 bridge those of low mass 
anemic spirals and lenticular galaxies. Lisker et al. (2006a) identify five of these 
galaxies as dwarf S0s (dS0s), noting that they differ from their more massive 
counterparts in terms of total luminosity and bulge size. The 
dS0 designation notwithstanding, images shown in 
Figure 7 of Lisker et al. (2006a) indicate that NGC 4305 and NGC 4620 have well-defined 
spiral structure, and only NGC 4306 and NGC 4497 lack spiral structure in 
the $g$-band images from Baillard et al. (2011). We note that spiral 
structure in intermediate mass disk galaxies in cluster environments 
may result from tidal forces (e.g. Kwak et al. 2019).
While not classified as a dwarf SO, Lisker et al. (2006a) 
conclude that NGC 4306 has a `certain disk', with integrated colors 
and a luminosity that are similar to those of the other five galaxies. 

\begin{table*}

\begin{center}
\begin{tabular}{cccccccccc}
\tableline\tableline
NGC & VCC & v$_r$\tablenotemark{a} & M$_K$\tablenotemark{b} & $(g-i)_{nuc}$\tablenotemark{c} & $(g-i)_{Tot}$\tablenotemark{d} & r$_e$\tablenotemark{e} & Morphology\tablenotemark{f} & n\tablenotemark{g} & $\Delta \alpha$\tablenotemark{h} \\
 & & (km/sec) & (mag) & (mag) & (mag) & (arcsec) & & & (arcmin) \\
\tableline
4305 & 0522 & 1888 & --20.7 & 0.76 & 0.99 & 29.1 & Sa & 0.6 & 133.5 \\
4306 & 0523 & 1981 & --20.0 & 0.74 & 1.00 & 16.9 & dSB0(s),N & 0.5 & 130.3 \\
4491 & 1326 & 497 & --21.1 & 0.33 & 0.99 & 25.3 & SBa(s) & 0.5 & 54.5 \\
4497 & 1368 & 1045 & --21.0 & 0.78 & 1.07 & 35.2 & SB0(s)/SBa & 0.7 & 47.2 \\
4584 & 1757 & 1779 & --20.0 & 0.30 & 0.96 & 19.6 & Sa(s) pec & 1.3 & 117.8 \\
4620 & 1902 & 1141 & --20.5 & 0.39 & 0.90 & 18.2 & S0/Sa & 1.5 & 166.8 \\
\tableline
\end{tabular}
\caption{Galaxy Properties}
\tablenotetext{a}{Radial velocity, from the NASA Extragalactic Database (NED)}
\tablenotetext{b}{Absolute $K$ magnitude, from the 2MASS Extended Object Catalogue, except 
for NGC 4491, which is from the 2MASS Large Galaxy Atlas (Jarrett et al. 2003). A 
distance modulus of 31.1 (Mei et al. 2007) is assumed.}
\tablenotetext{c}{Central color, computed from PSF magnitudes listed in the NED}
\tablenotetext{d}{Galaxy color, computed from total magnitudes listed in the NED}
\tablenotetext{e}{$H-$band half-light radius from McDonald et al. (2011)}
\tablenotetext{f}{From Binggeli et al. (1985)}
\tablenotetext{g}{$H-$band Sersic index from McDonald et al. (2009)}
\tablenotetext{h}{Angular offset from M87}
\end{center}
\end{table*}

	The spectra discussed by Davidge (2018a) indicate that most of these 
galaxies have luminosity-weighted central metallicities that are $\sim 1/2$ solar, 
and such a metallicity is more-or-less consistent with their luminosities. 
Gallazzi et al. (2005) examine the relation between stellar mass and metallicity 
using a large number of SDSS spectra. While there is considerable scatter in 
their metallicity vs. stellar mass relation, the ridgeline of the relation 
is consistent with a roughly half-solar mean metallicity for the 
galaxies in Table 1. A relation between metallicity and stellar mass/luminosity can 
also be constructed from gas phase metallicities, although the luminosity-weighted 
mean metallicity of stars that formed over a range of epochs and the chemical 
composition of the present-day gas may differ. The 
composite relation between [O/H] and M$_H$ presented by Saviane et al. (2008) in their 
Figure 9 suggests that a galaxy with M$_K \sim -20.5$ is expected to have log(O/H)+12 between 
8.3 and 8.4 if $H-K \sim 0.2$, or 0.3 -- 0.4 dex lower than solar. That the characteristic 
metallicities of these galaxies are consistent with their luminosities suggests that they 
have not been subjected to large-scale stripping of stellar material following their formation 
and early evolution, and that their dominant stellar 
components formed from material that was retained by the host galaxy 
during the epoch of disk assembly and early chemical evolution. The 
early evolution of these galaxies may then not have been influenced by processes 
that are usually associated with dense environments. 

	Most disk galaxies have radial spectrophotometric gradients that can be 
characterized in terms of metallicity and age. A metallicity gradient, in the sense of 
progressively lower luminosity-weighted metallicities towards larger radii, was found 
by Davidge (2018a) in five of the galaxies that they 
studied, and similar trends are common in disk galaxies (e.g. Goddard et 
al. 2017; Li et al. 2018). Simulations suggest that metallicity gradients in galaxies may 
have diverse origins (e.g. Pilkington et al. 2012), as metallicity at a given location 
in the disk can be influenced by factors such as the local mass density, 
the ability to retain gas if there are outflows powered by 
star formation and/or active nuclei, and the area within a galaxy where gas 
can be accreted -- either from a surrounding reservoir or from interactions with 
another galaxy -- and then cool sufficiently to form stars. These processes can also 
cause the radial distribution of metals to evolve with time (e.g. Magrini et al. 2016), 
as can interactions in a disk environment (e.g. Roskar et al. 
2008) or radial mixing that is driven by bars and mergers.

	The spectra discussed by Davidge (2018a) reveal older 
luminosity-weighted ages at larger radii. Such a radial age gradient is in the opposite 
sense to that seen in the vast majority of nearby isolated gas-rich disk galaxies 
(e.g. Gonzalez Delgado et al. 2017). While the age gradient in late-type disks 
is usually attributed to inside-out galaxy formation, the radial age distribution 
can be altered by secular (e.g. stellar migration or bar formation/collapse) or external 
(e.g. interactions with companions or an ambient intergalactic medium) processes that can 
affect the SFR and induce the radial mixing of stars. The 
red colors of the galaxies in Table 1 indicate that large scale 
disk star formation has been shut down for some time. The observed age gradients suggest that 
star formation did not end suddenly throughout these galaxies, but decreased over 
time scales of a Gyr or more, progressing from the low-density outer regions 
into the higher density central regions. While age gradients undoubtedly 
contain information imprinted in the course of the evolution of the host galaxy, the 
radial behaviour of luminosity-weighted age can be influenced by a number of processes. 
Age gradients on their own are then not unambiguous indicators of the process (or 
processes) that might have ended star formation in a cluster environment, such as 
strangulation of the gas supply, galaxy-galaxy encounters, or ram pressure stripping. 

	The galaxies in Table 1 have similar integrated colors at visible/red wavelengths, 
and in the absence of significant internal extinction the colors are indicative of light 
that is dominated by intermediate age or old populations. 
Despite having similar broad-band spectral energy distributions (SEDs) 
at visible and red wavelengths, there are galaxy-to-galaxy differences in 
the central colors, with NGC 4491, NGC 4584, and 
NGC 4620 having the bluest central colors. In fact, prominent emission lines are 
seen in the visible/red spectra of the centers of NGC 4491 and 
NGC 4584, while the central mid-infrared (MIR) colors of these 
galaxies point to significant amounts of hot dust (Davidge 2018a). That 
(1) there appears to be significant sources of ionizing radiation in the centers of two 
galaxies, and (2) there is a range of luminosity-weighted ages estimated for the disks 
from Balmer line depths (Davidge 2018a), suggests that while the galaxies in Table 1 share 
morphological similarities, they may be in different stages of evolution, and/or 
have been subjected to different evolutionary processes. 

\subsection{The Present Study}

	Light at wavelengths longward of $1\mu$m contains information for examining 
stellar content that is complementary to that obtained 
from light at shorter wavelengths. While main sequence turn-off (MSTO) and 
sub-giant branch (SGB) stars contribute significantly to the light at visible wavelengths, 
the light from intermediate age and old composite stellar systems at wavelengths longward of 
$1\mu$m is dominated by stars with low effective temperatures (e.g. Davidge 1990; Maraston 
2005). The construction of well-populated stellar libraries (e.g. Rayner et al. 
2009), have allowed comprehensive grids of model spectra to be 
computed that include the $1 - 2.5\mu$m wavelength region (e.g. Conroy \& 
van Dokkum 2012; Meneses-Goytia et al. 2015; Rock et al. 2016; Conroy et al. 2018). Not 
only do these models facilitate studies of the most evolved stars in stellar 
systems, but they have also fueled efforts to probe the mass function in the 
low stellar mass regime. Recent studies reveal a correlation between the slope of the 
mass function at the low mass end and the stellar velocity dispersion, in the 
sense of a steepening mass functions at low masses towards higher velocity 
dispersions (e.g. Rosani et al. 2018, Spiniello et al. 2014, Conroy 
et al. 2013, Cappellari et al. 2013, Ferreras et al. 2013).
Integrated light studies of the mass function are also important 
for very nearby objects (e.g. star clusters in the Galaxy) because 
they provide a complete census of low mass stars, 
whereas mass functions that are based on star counts may miss low mass 
stars that are in close binary systems. 

	In the current paper, long slit NIR spectroscopic observations 
of the galaxies in Table 1 are discussed. The spectra were recorded with the 
Flamingos-2 (F2) imaging spectrograph on Gemini South (GS) and cover 
wavelengths from $1.5 - 2.4\mu$m. The paper is structured as follows. The observations and 
the procedures used to remove instrumental and atmospheric signatures from the raw data 
are summarized in Section 2. Properties of the spectra are discussed in 
Sections 3 and 4, while comparisons with model spectra are the subject of Section 5. 
Radial variations in the spectra are examined in Section 6. 
The paper closes with a summary and discussion of the results in Section 7. 
 
\section{OBSERVATIONS \& REDUCTIONS}

\subsection{Description of the Observations}

	The spectra were recorded with the cryogenically-cooled F2 imaging 
spectrograph (Eikenberry et al. 2004) for programs GS-2016A-Q-84 and GS-2017A-Q-81 (PI 
Davidge). The detector in F2 is a Teledyne Hawaii-2 $2048 \times 2048$ array, with 
each pixel subtending 0.18 arcsec along the spatial direction. The spectra were 
recorded through a 4.5 arcmin long by 1 arcsec wide slit. Light was dispersed 
with the HK grism, with useable wavelength coverage extending from $1.5\mu$m to $2.4\mu$m. 
The spectroscopic resolution estimated from the widths of arc emission lines 
is $\frac{\lambda}{\Delta \lambda} \sim 400$ FWHM, 
which is consistent with published values \footnote[4]
{www.gemini.edu/sciops/instruments/flamingos2/spectroscopy/grisms}. This resolution 
is sufficient to recover age and chemical mixture at visible wavelengths (e.g. 
Choi et al. 2014).

	The spectra were recorded using conventional observing procedures for 
NIR spectra, with the galaxies offset in an `ABBA' nodding pattern along the slit, 
where `A' and `B' are different locations on the slit that 
were typically separated by $\sim 50$ arcsec. 
This offset was selected based on the angular extent of the 
galaxies and the need to retain guide stars to prevent time loss 
due to re-acquisition. In most cases this offset is at least $2\times$ the 
half light radius of a galaxy, and the slit was typically positioned at 
an angle that was close to the minor axis. An upper limit to 
the contamination from the `other' nod position in the outermost radial extraction 
interval is no more than 15\% based on the surface brightness profiles 
of low mass disk galaxies in Virgo discussed by McDonald et al. (2011). The cross-talk between 
nod positions in thus expected to be modest in the angular intervals considered in this 
study. For some targets it was necessary to use an offset that placed the target galaxy 
out of the science field for either the A or B position to avoid thermal emission from 
the guide probe entering the slit. \footnote[5]{The On-Instrument 
Wave Front Sensor (OIWFS) is preferred for F2 observations 
as it does not block NIR light in the science path. However, 
the OIWFS was inoperable when these data were recorded, and so a Peripheral Wave 
Front Sensor (PWFS) was used for guiding. The PWFSs are not part of F2, are not 
cooled, and are not in the instrumental focal plane. As a result, 
they can physically obstruct or cast shadows that occult the spectrograph slit, 
while also contributing to the thermal background.} 

	A 300 sec exposure was recorded at each slit location, 
and multiple ABBA cycles were repeated for each galaxy. A log of the observations 
can be found in Table 2. The last two columns of Table 2 show the estimated percentage 
$1\sigma$ noise levels per pixel in the central 2 arcsec (i.e. the Center and Region 
2 extraction areas, as defined in Table 3) of the $H$ and $K$ windows in the spectrum of 
each galaxy. Noise levels were measured near the peak overall throughput for each 
passband, and the spectra will be noisier near the edges of the $H$ and $K$ windows where 
telluric absorption lowers throughput. Not surprisingly, 
the two faintest galaxies (NGC 4306 and NGC 4584) have the noisiest spectra.

\begin{table*}

\begin{center}
\begin{tabular}{ccccc}
\tableline\tableline
NGC & Date (UT) & Number of & Noise & Noise \\
 & & Exposures\tablenotemark{a} & (H)\tablenotemark{b} & (K)\tablenotemark{b}\\
\tableline
4305 & March 25, 2016 & 10 & $\pm 0.8\%$ & $\pm 0.8\%$ \\
 & March 25, 2017 & 20 & & \\
 & & & & \\
4306 & February 11, 2017 & 36 & $\pm 1.1\%$ & $\pm 1.4\%$ \\
 & March 8, 2017 & 4 & & \\
 & & & & \\
4491 & April 1, 2016 & 20 & $\pm 0.7\%$ & $\pm 0.6\%$ \\
 & April 8, 2016 & 20 & & \\
 & April 4, 2017 & 20 & & \\
 & & & & \\
4497 & February 15, 2017 & 20 & $\pm 0.6\%$ & $\pm 0.6\%$ \\
 & April 5, 2017 & 20 & & \\
 & & & & \\
4584 & March 27, 2016 & 10 & $\pm 1.4\%$ & $\pm 1.9\%$ \\
 & March 30, 2017 & 20 & & \\
 & & & & \\
4620 & March 29, 2017 & 20 & $\pm 0.9\%$ & $\pm 0.7\%$ \\
 & April 2, 2017 & 20 & & \\
 & & & & \\
\tableline
\end{tabular}
\end{center}
\caption{Summary of Observations}
\tablenotetext{a}{Each exposure is 300 sec in duration.}
\tablenotetext{b}{$1\sigma$ percentage noise per pixel near the centers of the 
H and K passbands. The noise estimates apply to spectra within 2 arcsec of the galaxy centers.}
\end{table*}

	The observing strategy of alternating observations between two 
locations on the slit leads to a simple and efficient means of removing the 
`sky', which is dominated by telluric emission lines and the thermal background. The 
simplicity in sky removal occurs because the subtraction of two spectra taken 
at different slit locations results in the removal of background light that is not 
affected by fringing or non-uniformities in slit illumination. As for efficiency, 
science data are recorded at the same time as the background signal when this 
observing strategy is employed. While there is a time lag between when the galaxy and 
the background are observed at a given slit position, experience has shown 
that only minor variations in the amplitude of sky emission lines occur over time scales 
of a few hundred seconds during photometric or broken overcast observing conditions at GS.

	The spectrum of an early-type star was 
typically recorded before and after each block of galaxy observations to monitor 
telluric absorption. In some cases only one such `telluric standard' observation 
was recorded due to time limitations or the premature termination of observations 
to accomodate a change in observing conditions. These telluric star spectra 
are not suitable for absolute flux calibration because (1) they were often recorded 
during non-photometric conditions, and (2) there may be significant 
light loss outside of the slit.

	Spectra of dispersed light from a continuum lamp and an Ar arc that are 
in the Gemini facility calibration unit (GCAL) were also recorded. The former were 
recorded at various times on nights when the galaxies were observed, and these 
exposures were used to correct for non-uniform slit illumination and pixel-to-pixel 
variations in sensitivity (`flat-fielding'). The arcs were usually recorded 
at the end of an observing sequence.

\subsection{Data Reduction}

	The extraction of scientifically useable spectra 
involved a number of steps. The first was the subtraction of pairs of 
sequential offset exposures (i.e. subtracting galaxy frames 
taken at the B slit position from those at the A position, and visa versa). This 
removes additive components such as the background sky as well as artifacts 
that are introduced by the observing system and 
are common to both exposures. The latter includes 
the detector dark current as well as thermal signatures that originate in 
the instrument, the telescope, and from warm objects along the line of sight, such as 
dust on the F2 cryostat window.

	The next step was to divide the differenced spectra by the flat-field frames 
that were constructed from the GCAL continuum lamp observations. 
This step balances pixel-to-pixel variations in detector quantum efficiency, 
while also compensating for non-uniform illumination along the slit. A correction was 
then applied to remove optical distortions that 
curve the spectra perpendicular to the dispersion axis 
(the so-called spectrograph `smile'). The positions of 
emission lines in the arc spectra were traced at different locations along the slit, 
and a two-dimensional analytic function was fit to the resulting grid. 
A rectification function could then be generated and applied to the data
using this analytic representation of line curvature.

	At this point the processed data are a series of 
background-subtracted, flat-fielded, geometrically rectified images, most of which 
contain positive and negative spectra due to the subtraction of images performed
in the first step. Spectra within $\pm 20$ arcsec of each galaxy center were extracted 
for subsequent processing, and the positive and negative versions of these were 
co-added by subtracting the negative spectrum from its positive counterpart. All of the 
spectra for each galaxy were then combined to produce a final extracted 
two-dimensional spectrum. 

	Telluric absorption features were removed by 
dividing the galaxy spectra by a telluric reference spectrum, which typically was 
constructed from spectra of more than one star (see below). While the telluric 
spectra are not suitable for absolute flux calibration (Section 2.1), they 
still track large-scale wavelength-dependent variations in instrument response in a 
relative sense. This step thus removes structure 
from the spectra that is due to variations in instrumental 
response, simplifying the task of identifying a pseudo-continuum. 
The stars that monitor telluric features are hot sub-dwarfs, and some of the spectra 
contain Brackett series hydrogen lines. These lines
were fit with Gaussian profiles, and the resulting fits were subtracted from the 
stellar spectra prior to using them to correct for telluric absorption.

	The quality of telluric feature suppression was judged on a spectrum-by-spectrum 
basis via visual inspection, using the suppression of large-scale variations in 
wavelength due to telluric water absorption near $1.9\mu$m as the primary criterion. 
The center of this telluric H$_2$O band is heavily obscured, and thus
provides an extreme test for the correction of telluric features. 
Whenever possible, the average spectra of telluric stars 
that were observed before and after each observing block were used to remove 
telluric features. This resulted in an acceptable suppression of telluric 
features in most cases, while also boosting the S/N ratio 
of the telluric correction at the longest wavelengths in 
the $K$ atmospheric window, where the signal in the telluric spectra tends 
to be low. In the few cases when the suppression of telluric lines with the 
average telluric spectrum was found to be unsatisfactory 
then telluric corrections were made using individual stellar spectra, and the 
stellar template that best suppressed telluric features was adopted. 

	Interpolation between spectra in the $H$ and $K$ windows indicated that light in 
the telluric $1.9\mu$m feature was typically recovered at the $\pm 10\%$ level 
when averaged over many hundredths of a micron in wavelength. Averaging the signal 
over this wavelength interval was necessary to boost the S/N ratio in this 
region of low atmospheric transmission. The C$_2$ band near rest frame $1.76\mu$m 
falls in the shoulder of the atmospheric $1.9\mu$m H$_2$O band, and this features 
is of particular astrophysical interest (Sections 4 -- 6). The noise levels 
due to telluric features are up to $\pm 10\%$ per pixel near $1.8\mu$m (e.g. Figure 1).

	The penultimate step was wavelength calibration. Bright, 
isolated emission lines in the arc spectra were used as wavelength calibrators. 
Wavelength calibrated spectra were constructed with linear and logarithmic 
wavelength increments. The wavelength calibrated spectra were shifted into the 
restframe using radial velocities from the NASA Extragalactic 
Database (NED) \footnote[6]{https//ned.ipac.caltech.edu/}. These velocities were 
adopted over those measured directly from the data given the 
low spectral resolution of the F2 spectra. The veracity of the 
velocity corrections is evident in the galaxy-to-galaxy wavelength consistency of the 
spectroscopic features and the excellent agreement with features in model spectra. This is 
demonstrated in Figures 5 and 6, most noteably with the CO(2,0) band.

	Spectra were then extracted in the radial intervals listed in Table 3. Spectra 
in the same radial interval on both sides of the galaxy center were combined. 
The angular extent of the central region is based on the typical image quality, while the 
other binning intervals were selected to maintain more-or-less the same 
signal level as in the central spectrum. Still, the S/N ratio in most of the spectra 
plummets in Regions 3 -- 5, and so spectra in Regions 3 -- 5 are 
only considered for NGC 4491 and NGC 4497 (Section 6.2). The extracted spectra 
were normalized to a pseudo-continuum, which was found by fitting a low order 
polynomial to each spectrum after applying an iterative rejection filter to 
suppress absorption features. Such normalization avoids inherent uncertainties that 
may affect flux calibrated spectra (e.g. Conroy et al. 2018).

\begin{table*}
\begin{center}
\begin{tabular}{lc}
\tableline\tableline
Interval & Angular Range \\
Name & (arcsec) \\
\tableline
Center & 0 -- 0.75 \\
Region 2 & 0.75 -- 2.1 \\
Region 3 & 2.1 -- 5.1 \\
Region 4 & 5.1 -- 11.1 \\
Region 5 & 11.1 -- 23.1 \\
\tableline
\end{tabular}
\end{center}
\caption{Extraction Intervals}
\end{table*}

\section{VELOCITY MEASUREMENTS AND SPECTRAL RESOLUTION}

	The shape of absorption and emission 
features in a galaxy spectrum depends in large part on the spectroscopic resolution, 
which is defined by instrumental parameters (e.g. the ruling density on the 
dispersive element, slit width, and the imaging characteristics of the optics) 
coupled with the motions of stars and gas within the galaxy. An understanding of 
the spectral resolution is critical for determining population properties 
such as luminosity-weighted age and metallicity through comparisons with models. While 
the instrumental contribution to spectroscopic resolution can be measured directly from 
the widths of arc or telluric emission features, the contribution made by stellar 
motions within these galaxies must be assessed on a target-by-target basis. 

	Velocity dispersion measurements were made from the galaxy spectra with the FXCOR 
routine in IRAF. \footnote[7]{IRAF is distributed by the National 
Optical Astronomy Observatory, which is operated by the Association of Universities for 
Research in Astronomy (AURA) under cooperative agreement with the National Science 
Foundation.} The $2.15 - 2.3\mu$m wavelength interval, which 
contains the first overtone (2,0) CO band head, CN bands, and lines of 
Ca, Si, and Mg was adopted for the analysis. While the (3,1) CO 
transition is sampled with these data, the inclusion of this feature introduced 
side lobes in the correlation function that complicated efforts to 
measure a characteristic width. 

	K giant spectra from the library described by Rayner et al. (2009) were used 
as reference spectra for the velocity measurements. These stars were selected because of 
the spectroscopic similarities between solar neighborhood K giants and the galaxy centers 
(Section 4). Spectra of selected K giants were downloaded from the Infrared Telescope 
Facility website \footnote[8]{http://irtfweb.ifa.hawaii.edu/~spex/IRTF\_Spectral\_Library/}, 
and the results were smoothed and re-sampled to match the resolution and wavelength 
sampling of the F2 spectra. The correlation 
functions produced with these reference spectra did not yield measureable 
velocity dispersions, even though the galaxy centers 
are where the velocity dispersion and the S/N ratio should be highest. 
These results are thus consistent with the spectral resolution of all of the spectra 
being dominated by instrumentation, and so no adjustment to spectral resolution is made 
in this study for velocity dispersion.

\section{ABSORPTION AND EMISSION LINES IN THE CENTRAL SPECTRA}

	The examination of the spectra begins with a brief discussion 
of the features in the central spectra, while comparisons 
are made with model spectra in Section 5 and radial trends are examined in Section 6. 
The S/N ratio of the spectra is highest near the galaxy centers, and so 
the central spectra serve as good starting points for evaluating the 
stellar contents of these galaxies. Spectra of the central regions 
are compared in Figures 1 (H-band) and 2 (K-band), where prominent atomic and 
molecular features are identified. The atomic line and molecular band 
identifications in these figures are based on those made by Rayner et al. 
(2009) in figures throughout their paper, combined with the wavelength lists in 
their Tables 6, 7, and 10. 

\begin{figure}
\figurenum{1}
\epsscale{1.0}
\plotone{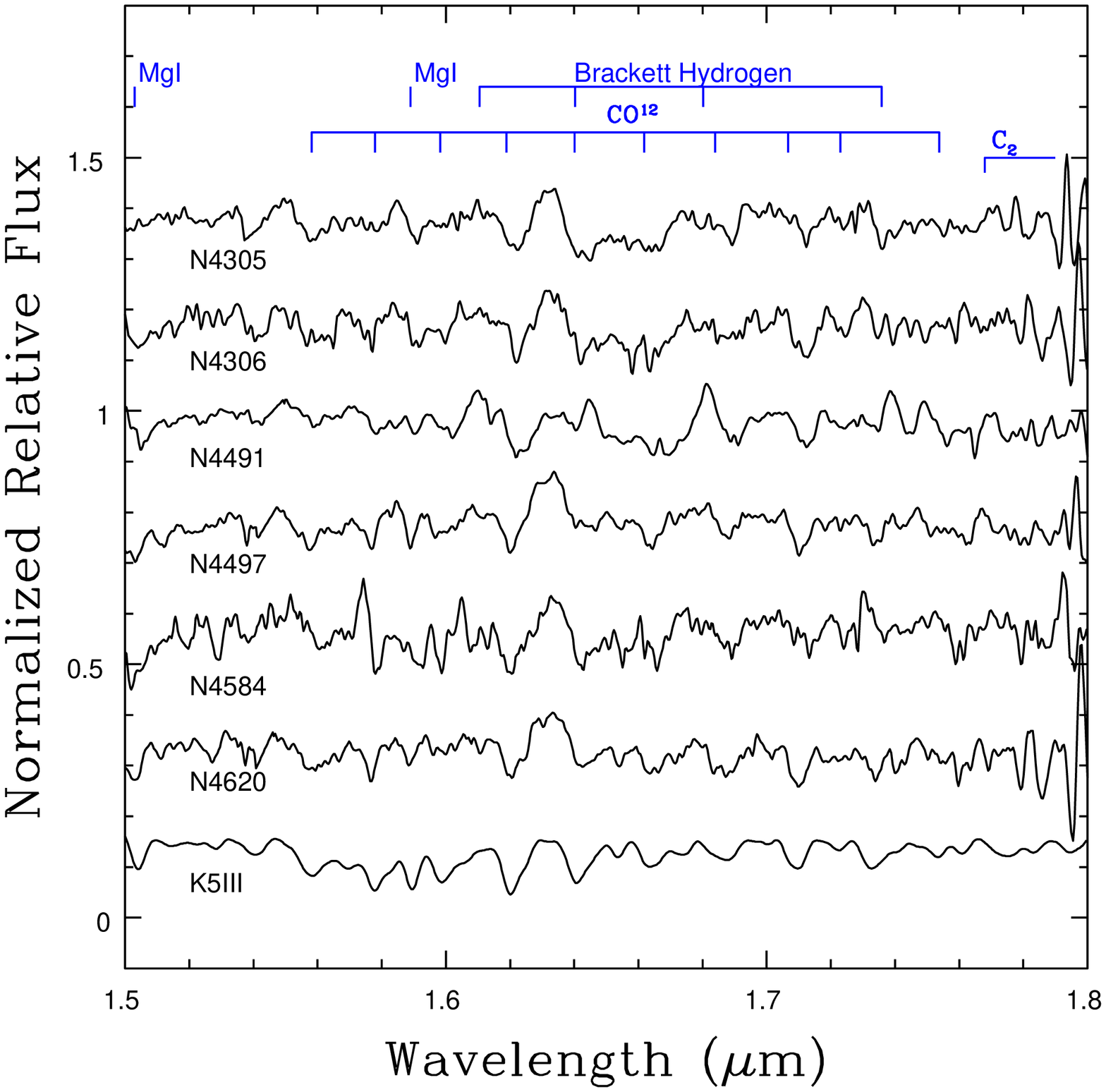}
\caption{$H-$band spectra of the central regions. The spectra have been normalized 
to a pseudo-continuum, and shifted vertically for the purposes of display. 
A spectrum of HD181596 (spectral-type K5III) is also shown. The HD181596 spectrum is 
from the Rayner et al. (2009) library, and was processed to 
match the spectral resolution and wavelength sampling of the F2 observations. 
The deepest absorption features in the galaxy spectra 
are the second overtone CO bands, while MgI absorption is 
also evident in most of the spectra. The wavelength interval longward of 
$1.76\mu$m contains the Ballick-Ramsey C$_2$ band, and in Section 5 it is 
shown that the spectra in this wavelength region tend to match that in the 
model spectrum of a moderately metal-poor 2 Gyr simple stellar population. 
The NGC 4491 spectrum contains [FeII] $1.644\mu$m 
emission, and this feature becomes more obvious when the 
stellar continuum is removed (Figure 8). Emission 
lines in the Brackett series are also present in that spectrum.}
\end{figure}

\begin{figure}
\figurenum{2}
\epsscale{1.0}
\plotone{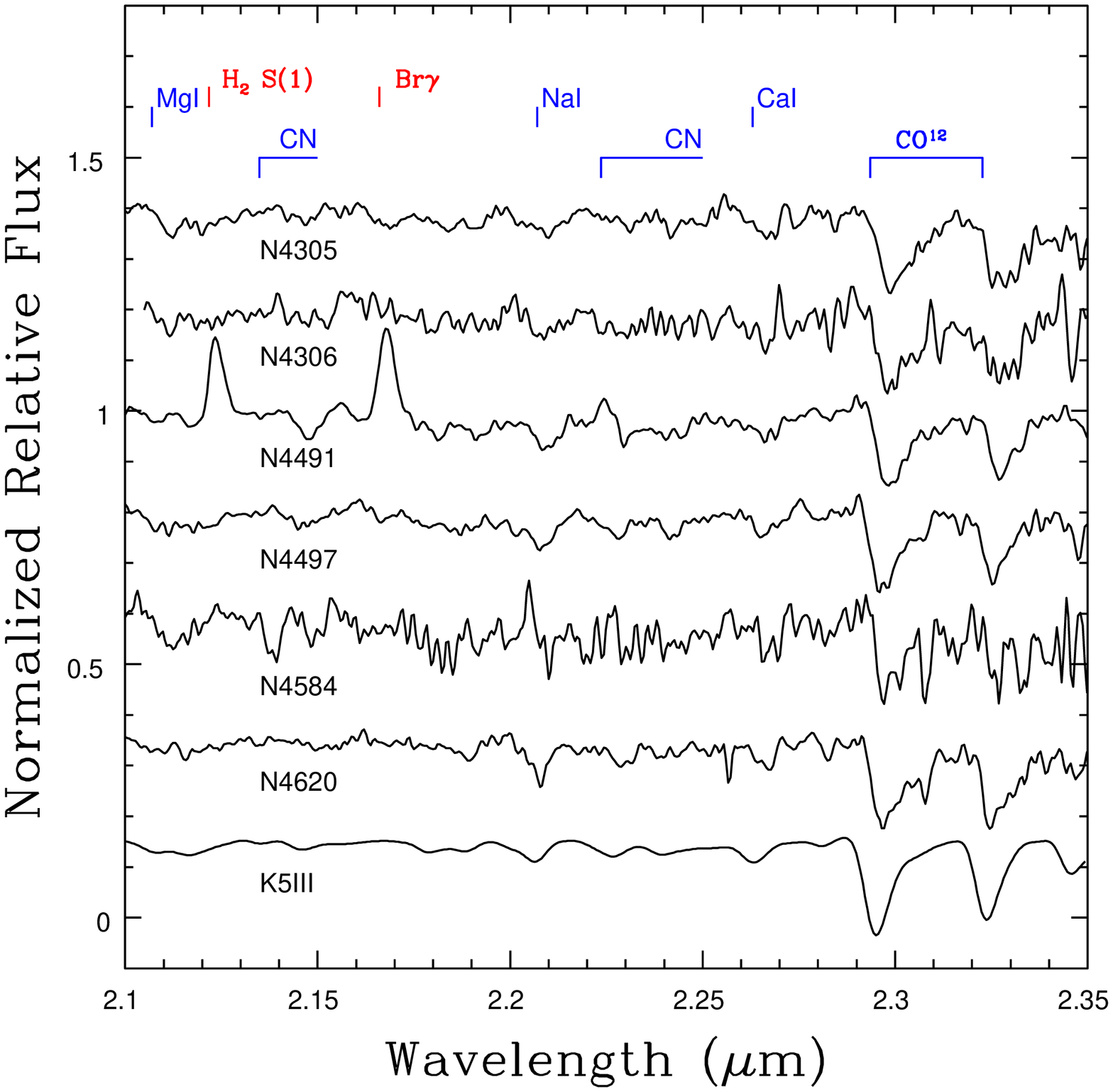}
\caption{Same as Figure 1, but showing $K-$band spectra. 
The first overtone CO bands are the dominant absorption features in the 
spectra, and these have depths that are similar to those 
in the K5III spectrum. Transitions involving 
Mg, Na, and Ca are present, and galaxy-to-galaxy 
variations in the depths of the NaI$2.21\mu$m doublet are evident. Br$\gamma$ and H$_2$ 
1--0 S(1) emission lines are seen in the NGC 4491 spectrum. However, emission lines are 
not evident in the NGC 4584 spectrum, even though the spectrum 
of the central regions of that galaxy at visible wavelengths
contain emission features. This suggests a source for ionizing radiation 
in NGC 4584 that is distinct from that in NGC 4491 (see text).}
\end{figure}

	The spectrum of the K5III star HD181596 is shown in Figures 1 and 2 to 
aid in the identification of blended atomic and molecular lines. 
This star was selected as a reference because 
its spectrum shares features that are similar to (but not identical to) 
those in the galaxy spectra. The spectroscopic properties of HD181596 are those 
of an evolved star that has a more-or-less solar metallicity and chemical mixture. Its 
spectrum in Figures 1 and 2 is from the Rayner et al. (2009) library, 
and has been smoothed and re-sampled to match the spectral resolution and wavelength 
sampling of the F2 spectra. The widths of features in the smoothed HD181596 
spectrum match those in the F2 spectra, although noise in the latter make some features 
appear sharper than in the former. Like the galaxy spectra, the HD181596 spectrum 
was normalized to a pseudo-continuum that was found by fitting a low order 
function to pseudo-continuum points. 

	The dominant absorption features in Figure 1 are the second-overtone 
bands of CO. The modest spectral resolution of the F2 spectra complicates 
the detection of atomic absorption features, although absorption lines 
of MgI are seen in most of the $H-$band spectra. Emission lines of the hydrogen 
Brackett series are present in the NGC 4491 spectrum, and the presence of these lines 
is not surprising given the line emission in the visible spectrum of that galaxy 
found by Davidge (2018a). [FeII]$1.64\mu$m emission is also seen (Section 6). 
However, obvious emission features are not evident in the NGC 4584 
spectrum, even though the visible/red spectrum of that galaxy has 
H$\alpha$ and [SII] in emission, with equivalent widths 
that are only slightly weaker than those in NGC 4491. 

	The characteristic age and metallicity estimates from Figures 16 and 17 
of Davidge (2018a) are such that some of these galaxies might be expected to have high 
C star frequencies. The Ballick-Ramsey C$_2$ band at $1.76\mu$m 
can be a prominent feature in the spectra of late-type C stars, and has been detected 
in the integrated light of nearby galaxies with large intermediate 
age populations (Miner et al. 2011; Davidge 2015b, 2016). 
At low redshifts the C$_2$ band falls in the shoulder of the 
telluric H$_2$O band centered near $1.9\mu$m, and the recession velocities of 
galaxies in Virgo push the C$_2$ band into this water band. 
The rest frame galaxy light near $1.76\mu$m thus has diminished throughput 
and is subject to telluric absorption features from H$_2$O transitions. 
The problems associated with telluric water absorption notwithstanding, 
there is a tendency for the mean signal in Figure 1 to be depressed 
in the NGC 4305, 4306, and 4620 spectra at wavelengths where C$_2$ absorption might 
be expected. In Section 5 it is shown that the spectra at these wavelengths are 
consistent with models of a moderately metal-poor simple stellar population (SSP) with an 
age of 2 Gyr.

	The first overtone bands of CO are the deepest features 
in Figure 2. There are other molecular signatures in 
Figure 2, and some of the absorption features at wavelengths immediately longward 
of $2.13\mu$m and $2.23\mu$m are due to CN. These CN bands are not sensitive 
indicators of C stars (e.g. Figure 34 of Rayner et al. 2009).

	The Ca triplet near $2.26\mu$m is detected in all of the 
spectra, as is the NaI doublet near $2.21\mu$m. 
Individual lines of NaI and CaI within these features are not resolved, 
and so they appear as broad absorption features.
There are galaxy-to-galaxy differences in the depth of 
the NaI doublet, and these are of interest as NaI$2.21\mu$m 
in some early-type galaxy spectra has been found to be deeper than predicted 
by models that assume a solar chemical mixture 
and solar neighborhood mass function. Rock et al. (2017) attribute stronger-than-expected 
NaI$2.21\mu$m absorption to a super-solar [Na/Fe] combined with a bottom-heavy mass 
function. That there are galaxy-to-galaxy differences in the depth of NaI$2.21\mu$m 
thus suggests that there may be differences in stellar content among the Virgo galaxies 
that can not be attributed to the dispersion in luminosity-weighted age and/or 
metallicity that might be expected among a homogeneous sample of objects 
in a cluster environment that are at different stages of evolution. Possible origins of 
the galaxy-to-galaxy NaI$2.21\mu$m differences are discussed in Section 7.

	The $K-$band spectrum of NGC 4491 differs from 
the others in that there are molecular and atomic Hydrogen 
emission lines. The relative strengths of the H$_2$ S(1) and Br$\gamma$ lines 
yield insights into the physical conditions in the emission region. The ratio of the 
strengths of these two lines is $\frac{H_2}{Br\gamma} \sim 0.9$, and this is near the 
high end of the range seen among star-forming galaxies (e.g. Puxley et al. 1990).

	Veiling of stellar spectra by 
nebular emission affects the depths of absorption features, 
with the NIR spectra of the central regions of NGC 253 being one example (Davidge 2016).
However, it is unlikely that veiling affects the depths of absorption lines 
in the central spectrum of NGC 4491. If a significant emission component were present 
then the CO bands in NGC 4491 might be expected to deepen with progressively larger 
distances from the galaxy center as the fractional contribution from emission dimishes, 
and such a trend is not seen (Section 6). Nebular emission dominates 
the light output near $2\mu$m in SSPs with ages of 1 Myr, but by an age of 5 Myr the 
emission component becomes substantially weaker (e.g. Figure 11 of Byler et al. 2017). 
If there is continuous star formation then an equilibrium state is reached 
such that relative line strengths in the emission spectrum are 
similar to those in a system with an age of a few Myr (Byler et al. 2017). 
High levels of nebular continuum emission are thus only expected if the center of NGC 4491 
is viewed during a relatively short window in time after the onset of star formation. 

\section{COMPARISONS WITH MODELS}

	Comparisons are made with model spectra from the E-MILES suite (e.g. Rock et 
al. 2016) in this section. These models span a broad range of metallicities and ages, 
with wavelength coverage that extends from the near-UV to the MIR. 
The models use the stellar library discussed by Rayner et al. (2009), which 
is made up largely of bright Galactic stars. Hence, the models are most applicable 
for stellar systems that have near-solar metallicities and chemical mixtures. 
Model spectra in the E-MILES compilation have been constructed using the Padova00 (Girardi 
et al. 2000) and BaSTI (Pietrinferni et al. 2004; Cordier et al. 2007) isochrones. 
In the current paper, comparisons are made with models that are based 
on the BaSTI isochrones. A Chabrier (2001) mass function is assumed.

	Conventional practices for probing stellar content are to examine 
spectral indices and/or directly fit models to the observed spectra. However, a 
different approach is adopted here in an effort to highlight subtle changes in the 
spectra: model spectra are subtracted from a reference spectrum and the residuals 
are examined. Systematic artifacts due to uncertainties in the removal of telluric 
absorption and emission features are suppressed when 
such `differenced spectra' are considered. The suppresion of these features are 
of particular importance when probing the wavelength region that contains 
the C$_2$ band at $1.76\mu$m in the Virgo galaxy spectra, as it is in the shoulder 
of a deep telluric H$_2$O band..

	Model spectra were downloaded from the MILES website 
\footnote[9]{http://www.miles.iac.es}, and these were
smoothed and re-sampled to match the spectral resolution and sampling 
of the F2 spectra. The sensitivity of spectral features to changes in age and [M/H] 
(the abundance of all metals with respect to the solar 
value) depend on spectral resolution, in the sense that 
higher resolution spectra have the potential to yield more information and tighter 
constraints on relevant parameters (e.g. Dahmer-Hahn et al. 2018). This being said, 
ages, metallicities, and chemical abundances may be deduced from visible spectra with 
resolutions as low as $\frac{\lambda}{\Delta \lambda} \sim 150$ 
(e.g. Appendix 2 of Choi et al. 2014), although the bright main sequence stars 
that contribute significant age and metallicity information at visible 
wavelengths make a much smaller contribution to the integrated light in the NIR. Thus, 
changes in features in the NIR in response to variations in age and metallicity 
are more subtle.

\subsection{Trends in Model Spectra with $\frac{\lambda}{\Delta \lambda} = 400$}

	The response of model spectra with the spectral resolution of 
the F2 observations to changes in age and metallicity is examined in 
Figures 3 and 4. Emphasis is given to features that are of interest 
for examining the stellar contents of old and intermediate age populations. Each figure 
shows selected wavelength intervals from a reference spectrum of a 2 Gyr SSP with [M/H] 
=--0.25, as well as differenced spectra that are the result of subtracting models with 
different metallicities and ages from the reference spectrum. The 
age and metallicity of the reference spectrum is typical of that deduced from the 
visible-red spectra by Davidge (2018a). James \& Percival (2018) find similar 
luminosity-weighted ages in disks that have been swept by bars. The reference 
spectrum has been continuum-corrected and normalized to unity, and then 
shifted vertically in Figures 3 and 4 for display purposes. The 
differenced spectra were constructed by subtracting normalized, 
continuum-corrected model spectra, and these have also been shifted vertically 
for display purposes. CaI$2.26\mu$m is in a CN band, and the spectra of this feature have 
been adjusted to match a localized pseudo-continuum. Given that the differenced spectra were 
constructed from normalized spectra then the increments along the vertical axis in Figures 
3 and 4 directly measure fractional variations in the strengths 
of features as age and metallicity change; each 0.01 increment along the 
vertical axis corresponds to a 1\% difference between subtracted model spectra. 

	The left hand panel of each Figure examines the behaviour of 
a composite feature termed $\Sigma$CO. There are numerous second-overtone 
CO transitions in the $H-$band and these contain similar encoded information. 
$\Sigma$CO is an attempt to multiplex the information content of these 
features by combining them together. A complication is that the 
CO bands are blended with other features. Transitions were selected for combination 
using the K5III spectrum in Figure 1 as a guide to overcome potential problems 
with blending. Based on their shape and isolation from other obvious features, 
the (3,0),(6,3), (7,4), and (10,7) transitions were selected to construct $\Sigma$CO. 
A $0.01\mu$m wavelength interval centered on each transition was extracted, and these 
were then averaged together to produce $\Sigma$CO. Taking the median 
instead of the mean would suppress information from outliers; however, 
outliers are not expected to be present since the transitions used to construct 
$\Sigma$CO were selected based on their shape and isolation. Taking the mean also avoids 
problems due to uncertainties in the placement of the local pseudo-continuum near 
each transition.

\begin{figure}
\figurenum{3}
\epsscale{0.55}
\plotone{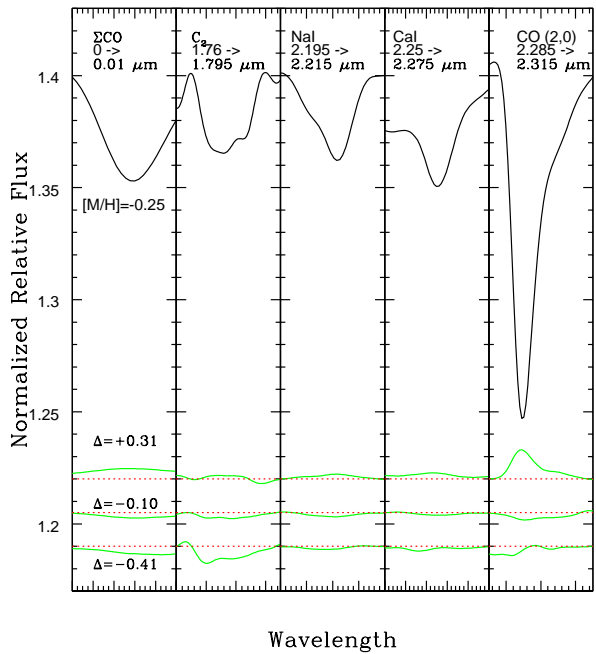}
\caption{The response of the E-MILES models to changes in metallicity. Each panel 
highlights a feature that probes stellar content. $\Sigma$CO is the composite second overtone 
CO band that is defined in the text. The wavelength intervals sampled are specified 
at the top of each panel - in the case of $\Sigma$CO a $0.01\mu$m 
wavelength interval is shown. The black lines show the 2 Gyr 
[M/H] = -0.25 reference spectrum, while the green lines show the differences between that 
spectrum and E-MILES models with [M/H] = +0.06 ($\Delta$[M/H]=+0.31), [M/H] = -0.35 
($\Delta$[M/H]=-0.1), and [M/H] = -0.66 ($\Delta$[M/H]=-0.41). 
The reference spectrum has been processed to have the same spectral resolution 
and wavelength sampling as the F2 spectra, then continuum-corrected and normalized to 
unity. The results have been shifted vertically for display purposes. The green lines were 
produced by subtracting model spectra that have been continuum-corrected and 
normalized to unity before being shifted vertically for display purposes. 
The red lines indicate the shifted zero levels in the differenced spectra. 
As the differenced spectra were constructed from spectra 
normalized to unity then the increments along the vertical axis 
directly gauge fractional variations in the differenced spectra. As explained 
in the text, the non-linear behaviour of CO(2,0) with metallicity may be due 
to the models adopted for evolution on the upper AGB. The C$_2$ band also changes with 
metallicity, but in a way that is perhaps unexpected -- the differenced spectra 
for [M/H] $= -0.66$ indicate a {\it shallower} C$_2$ band than in the [M/H] $= -0.25$ 
models even though the C star frequency increases. This is due to 
the bluer giant branch sequences that are expected 
as metallicity drops and age remains constant, although 
the sample of C stars in the stellar library may also play a role (see text). 
The other features are less sensitive to changes in [M/H].}
\end{figure}

\begin{figure}
\figurenum{4}
\epsscale{0.9}
\plotone{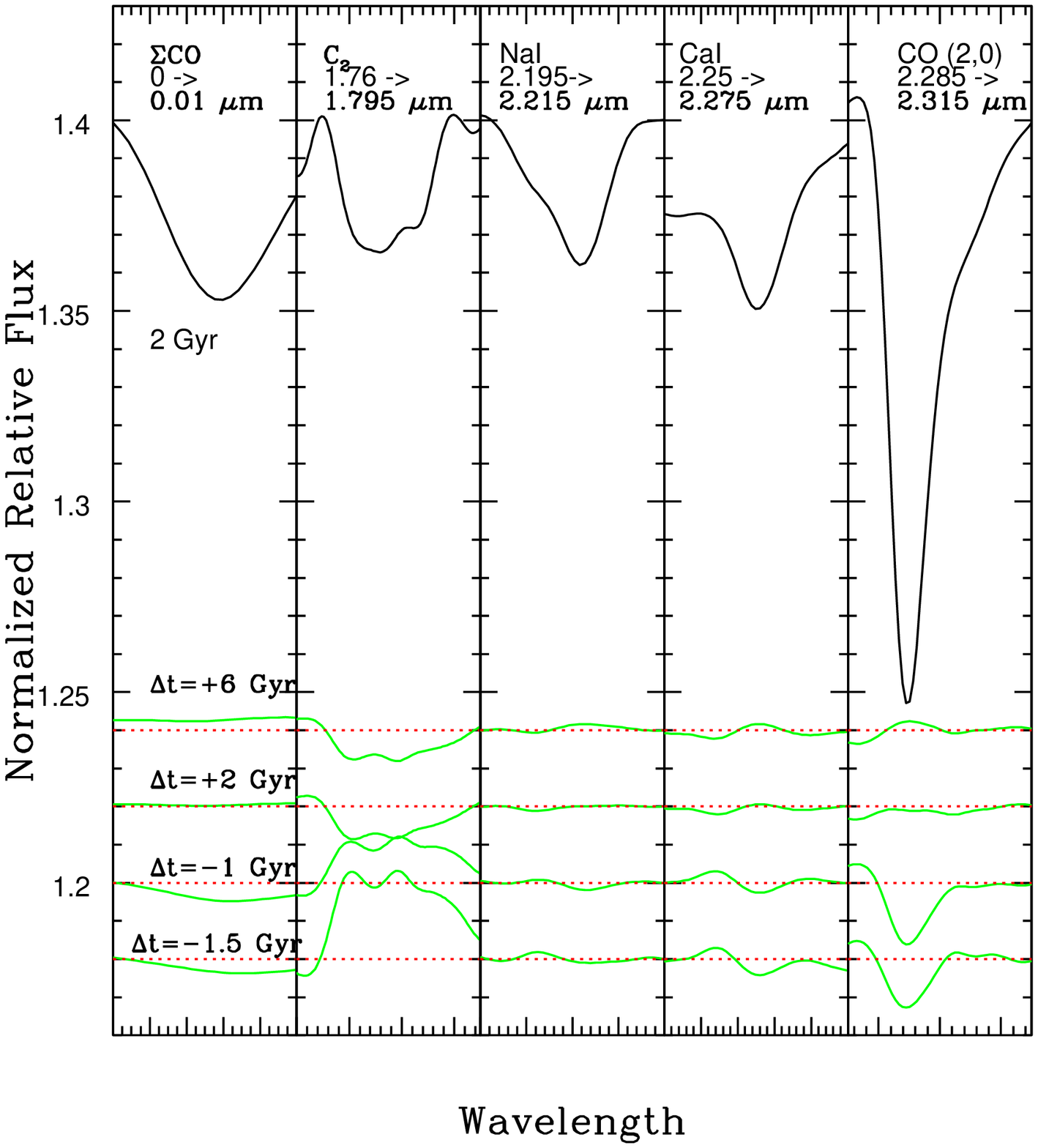}
\caption{Same as Figure 3, but examining sensitivity to age.
The green lines show the differences between the 2 Gyr reference spectrum and 
models with ages of 8 Gyr ($\Delta$t = 6 Gyr), 4 Gyr ($\Delta$t = 2 Gyr), 1 Gyr 
($\Delta$t=-1 Gyr) and 0.5 Gyr ($\Delta$t = -1.5 Gyr). All models have 
[M/H] = --0.25. The NaI$2.21\mu$m is not sensitive to changes in age, likely as it is 
strongest in the spectrum of very low mass stars. 
The C$_2$ band weakens towards progressively olders ages, reflecting changes in the 
contributions made by C stars in the models. The weakening of the CO bands for ages 
younger than 2 Gyr reflects the change in the color of the giant branch with age 
when metallicity is held constant.}
\end{figure}

	The depth of $\Sigma$CO changes in a similar manner to CO (2,0) in Figures 3 and 4, 
and so in the following discussion we focus our attention on CO(2,0). 
The sensitivity of the CO(2,0) band to [M/H] reflects changes in 
line strength due to chemical content, as well as changes in the temperature 
of the giant branch, which at a fixed age shifts to lower values as metallicity 
increases. The temperature dependence of the CO(2,0) band is such 
that shifting the temperature of the giant branch to cooler values 
at a fixed age causes the depth of the CO(2,0) band to increase, and 
this behaviour is seen in the [M/H] $= +0.31$ differenced spectrum. However, there are 
departures from these basic expectations. For example, 
the differenced spectra in Figure 3 indicate that the CO(2,0) 
band in the [M/H] $= -0.66$ model is slightly {\it stronger} than in the [M/H] 
$= -0.25$ model, which is contrary to what might be expected. 

	There are inherent uncertainties in the 
physics used to model highly evolved stars, and these propogate into 
the isochrones from which the model spectra are constructed. The depth of the CO(2,0) band 
in the [M/H] $= -0.66$ model might be a consequence of how 
the latest stages of AGB evolution are tracked in the models. 
The temperature of the AGB shifts to higher values 
at a fixed age as metallicity is lowered. However, evolution on the AGB is also 
sensitive to the rate of mass loss, in the sense that 
if the mass loss rates decrease towards lower metallicities then AGB stars 
with progressively lower metallicities will evolve to more advanced stages before 
leaving the AGB (e.g. Bowen \& Willson 1991; Willson 2000). Still, 
mass loss rates estimated for highly evolved stars in the SMC, LMC, and the Galaxy 
are very similar, even though these galaxies span $\sim 0.7$ dex in [M/H]
(e.g. Groenewegen \& Sloan 2018). A trend of higher AGB-tip 
luminosities towards lower metallicities is seen in the BaSTI isochrones, which adopt the 
mass loss formulism of Vassiliadis \& Wood (1993) for the AGB 
(Cordier et al. 2007). If the AGB-tip is allowed to evolve to lower surface gravities and 
higher luminosities at [M/H] $= -0.66$ when compared with [M/H] $= -0.25$ then the 
expected trend of decreasing CO(2,0) strength with decreasing metallicity in the 
models might be reversed, as is seen in Figure 3. 

	The C$_2$ band in the [M/H] $= -0.66$ model is weaker than in the 
[M/H] $= -0.25$ model. Naively, this seems to contradict the empirical 
relation between C star frequency and metallicity (e.g. Battinelli \& Demers 2005), 
which is such that the ratio of C to M stars increases in galaxies 
as metallicity drops. The contribution that C stars make to the NIR light, and hence 
the depths of signature C star features such as C$_2$, might then be expected to strengthen 
as metallicity is lowered. However, the temperature of the AGB sequence increases as 
metallicity drops, and so plays a role in defining the contribution that C stars 
make to the NIR light. Still, C star signatures strengthen as 
metallicity decreases in models discussed by Maraston (2005). 

	The sample of C stars that are in the model library and the contributions 
that they make to the NIR spectrum also influence the depth of C molecules in model
spectra. For example, gaps or omissions in the types of C stars in the library will 
compromise the ability of models to match the spectra of real systems. C stars in 
the model library may also have diverse spectroscopic properties, possibly related to 
particular circumstances that influenced their evolution, such as binarity. Figure 34 of 
Rayner et al. (2009) compares the spectroscopic properties of C stars in 
their library. The coolest C star in their sample is R Lep, and the C$_2$ band in 
the spectrum of that star is shallower than in the spectra of the two warmer C stars, 
even though the CO(2,0) bands in R Lep are deeper than in the spectra of the other stars. 
Varying the fraction of light that stars like R Lep contribute in the model will 
potentially affect the properties of the C$_2$ and CO(2,0) features in the model spectra 
of intermediate age populations in a complicated way.

	The comparisons in Figure 4 indicate that the depths of the C$_2$ and 
the CO(2,0) bands are sensitive to changes in age for systems with half-solar metallicities. 
NaI$2.21\mu$m is remarkably insensitive to age variations, and this is likely due 
to this feature being deepest in the spectra of stars that are on the lower main sequence. 
The C$_2$ band grows in strength in the models that are younger 
than 2 Gyr, reflecting the increase in the C star frequency in models with younger ages, 
coupled with the higher luminosity of the AGB-tip that occurs when progenitor 
mass increases. As for the CO(2,0) band, the changes in the depth of this feature
with age are due to the temperature of the giant branch. 
The temperature of the giant branch changes only slightly with age 
among systems with metallicities near those considered here and ages in 
excess of a few Gyr (e.g. Figure 3 of Pietrinferni et al. 2004), and so 
the depths of CO(2,0) in the 2 Gyr and 8 Gyr models are only modestly different. 
In contrast, the giant branch temperature changes with age at a more rapid pace in 
systems with ages younger than 3 Gyr. This results in a greater rate of change 
in the depth of the CO(2,0) band with age among systems that are younger than 2 Gyr 
than is the case among older systems.

\subsection{Comparing the F2 Spectra with Models}

	The comparisons in Figures 3 and 4 indicate that galaxy-to-galaxy variations 
in age and metallicity of the size found from the GMOS spectra 
will alter the depths of key NIR spectroscopic features in the F2 spectra by only a 
few percent. We thus limit comparisons with models to 
the central spectra, as this is where the S/N ratio is highest. 
The one exception is NGC 4491, where comparisons with models are made 
with the the Region 3 spectrum to avoid nebular emission near the center of that galaxy. 

	The goal of these comparisons is not to deduce an independent 
SFH. This would be a challenging task given the reduced contribution made by 
stars near the main sequence turn-off to NIR light and uncertainties in the models of 
stars in the advanced stages of evolution (e.g. Dahmer-Hahn et al. 
2018). In addition to uncertainties in the models, variability among 
the most evolved objects will affect their contribution to the total light 
(e.g. Davidge 2014). In the current paper we thus focus 
on determining if the NIR spectra are consistent with the luminosity-weighted 
ages and metallicities estimated from the GMOS spectra. Major discrepancies 
between models and NIR spectra will reveal uncertainties in the evolution of 
highly evolved stars and the assumptions used to construct the models, such as the 
nature of the mass function among low mass stars. Furthermore, while galaxies are complex 
stellar systems that contain stars spanning a range of ages and metallicities, the 
comparisons made here are restricted to models of SSPs, as such comparisons provide 
the luminosity-weighted ages and metallicities 
that will be compared with those found from the GMOS spectra.

	The H$\beta$ and Mg$_2$ indices measured by 
Davidge (2018a) indicate that the luminosity-weighted ages in 
the central regions of these galaxies fall between 1.5 and 3 Gyr. 
These same data also suggest that the luminosity-weighted [M/H] range from --0.3 to just 
above solar. Comparisons are thus made with the 2 Gyr [M/H] = --0.25 model spectrum, 
and the results are shown in Figures 5 (NGC 4305, NGC 4306, NGC 4491) and 6 
(NGC 4497, NGC 4584, NGC 4620). As with Figures 3 and 4, comparisons are restricted 
to specific features, and the impact of changes in age and/or metallicity 
can be assessed using the comparisons made in those figures. 
The $\Sigma$CO composite feature is not included in these comparisons 
as its response to changes in age and metallicity was shown previously 
to be similar to that of CO(2,0).

\begin{figure}
\figurenum{5}
\epsscale{0.95}
\plotone{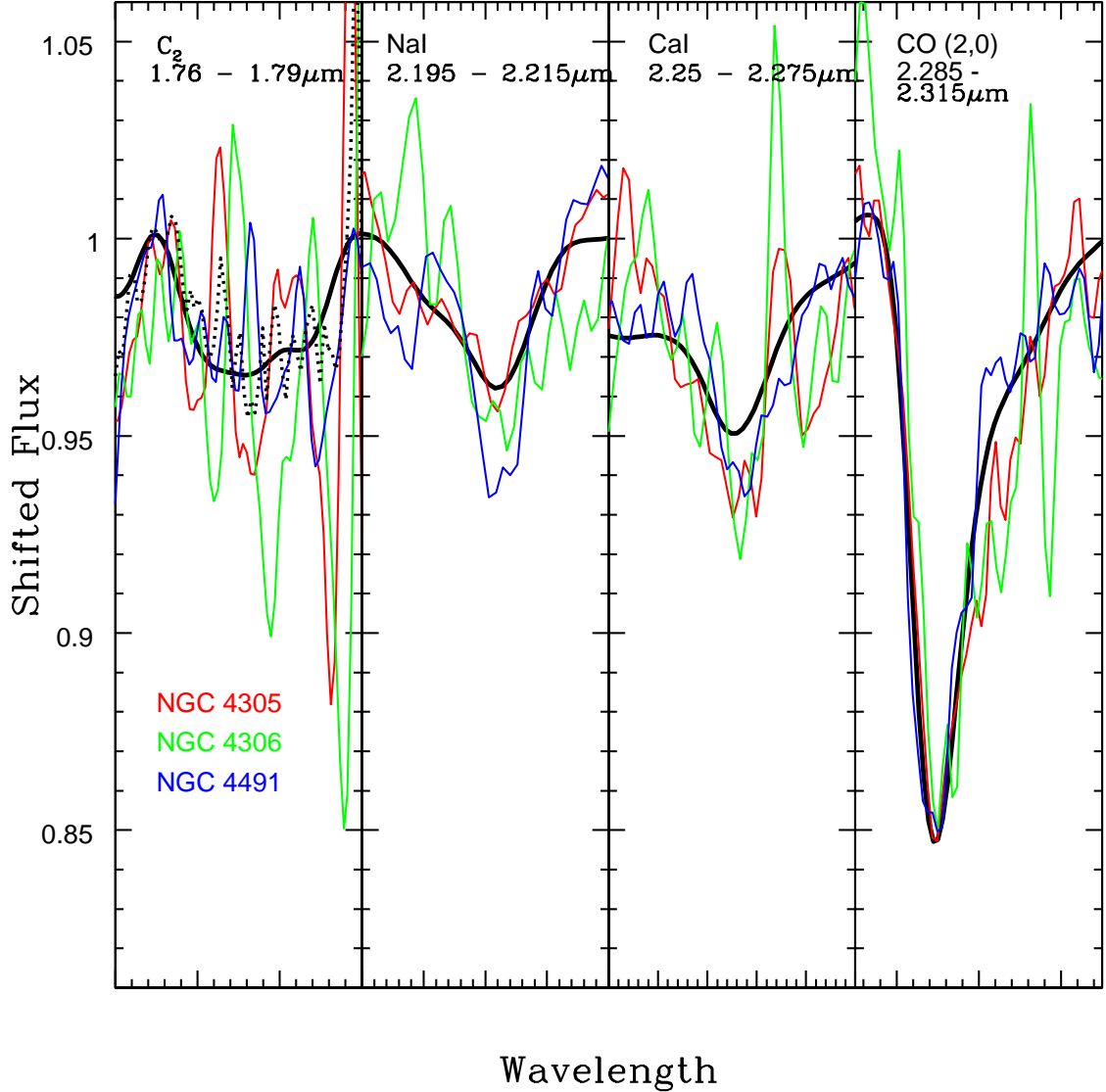}
\caption{Selected wavelength intervals in the 2 Gyr [M/H] = --0.25 model reference spectrum 
are compared with spectra of NGC 4305 center (red lines), NGC 4306 center (green lines), 
and NGC 4491 Region 3 (blue lines). The median of the continuum-corrected and 
normalized center spectra of all six galaxies is shown as a black dotted line 
in the left most panel. The C$_2$ band in NGC 4491 more-or-less matches the models at the 
$\pm 2\%$ level, whereas the noise in the NGC 4305 and NGC 4306 spectra makes comparisons 
with the model at these wavelengths problematic. However, 
the median spectrum (see text) agrees with the reference model 
at roughly the $\pm 1\%$ level. The NaI $2.21\mu$m line in NGC 4305 and NGC 4306 is 
well-matched by the model, but this feature is deeper in NGC 4491 than in the 
model spectrum. A galaxy-to-galaxy dispersion in CaI depths is not 
evident. However, CaI appears to be consistently deeper than predicted by the reference 
model. The depths of the CO(2,0) band show good galaxy-to-galaxy agreement.} 
\end{figure}

\begin{figure}
\figurenum{6}
\epsscale{1.0}
\plotone{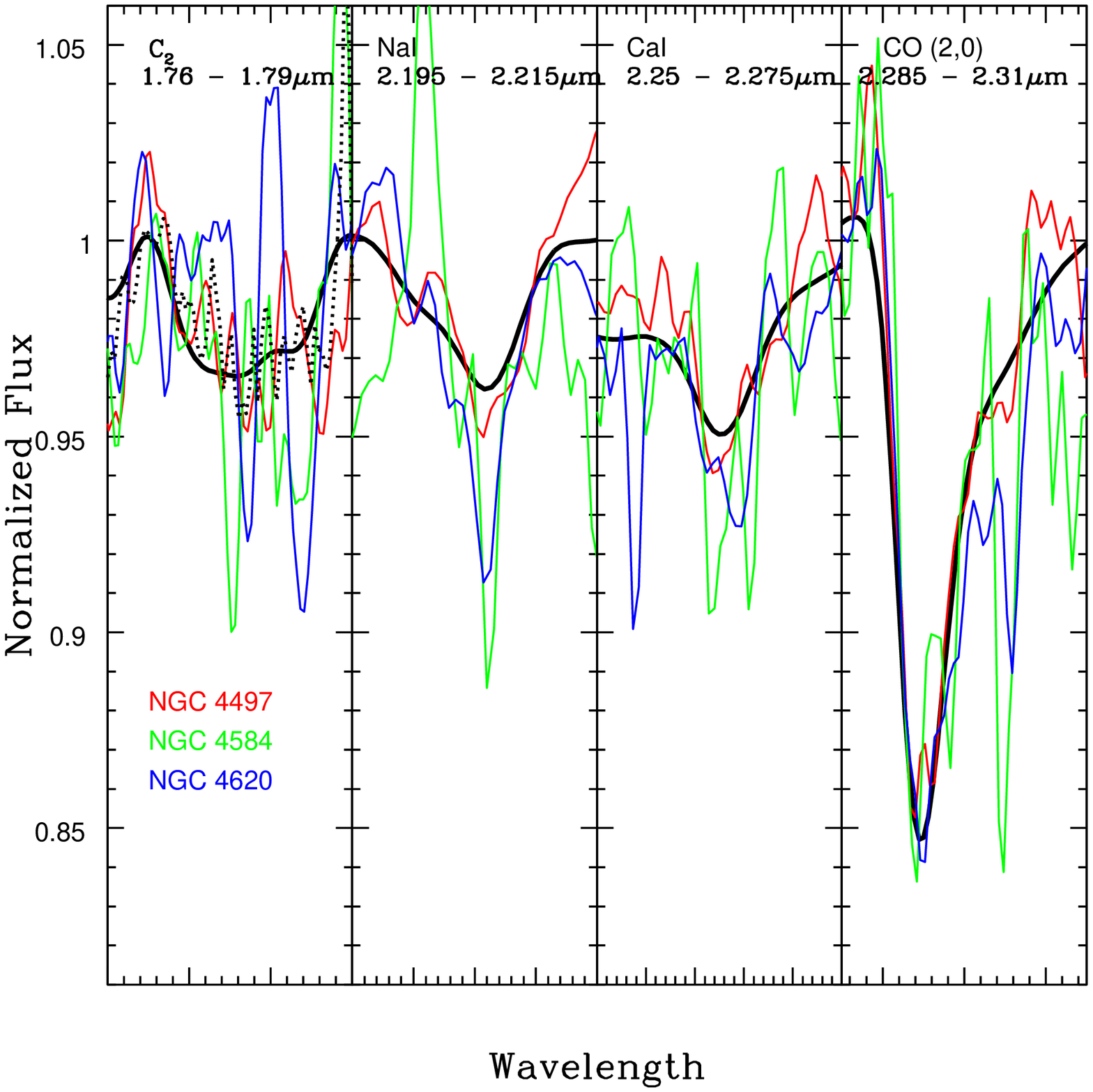}
\caption{Same as Figure 5, but showing the central spectrum of NGC 4497 (red lines), 
NGC 4584 (green lines), and NGC 4620 (blue lines). 
The NGC 4497 spectrum matches the reference model 
at wavelengths near the C$_2$ band at a level 
that is comparable to that of the median spectrum. 
The NaI$2.21\mu$m doublet in NGC 4584 and NGC 4620 is deeper than predicted 
by the model, although the NGC 4584 spectrum has a poor S/N ratio. 
There is galaxy-to-galaxy agreement with the CaI$2.26\mu$m triplet. 
The CO (2,0) features in all three galaxies more-or-less match the reference model.}
\end{figure}

	The depth of the CO(2,0) band in the spectra of most of the galaxies 
is consistent with that in the 2 Gyr [M/H] = --0.25 reference model. The agreement 
with the models in the right hand panels of Figures 5 and 6 is not perfect, as there 
is a tendency for the spectra of NGC 4305, NGC 4306, and NGC 4620 to fall below 
the reference model to the right of the band center in these figures. 
This wavelength region corresponds to the onset of telluric H$_2$O absorption, and is 
also where the throughput of the spectrograph and the F2 HK band-limiting filter plummets. 
The differences between the observations and models is likely due to difficulties 
removing telluric absorption and identifying the pseudo-continuum in the spectra at 
low S/N ratios. In any event, the difference between the models and the spectra of 
these galaxies to the right of the CO(2,0) band is at the 1 -- 2\% level, and so is 
subtle. NGC 4491 and NGC 4497 are the brightest galaxies in our sample, 
and their spectra have the highest S/N ratio. The sequences for 
these galaxies agree with the reference model spectrum near CO(2,0) in Figures 5 and 6.

	The comparisons in Figures 3 and 4 indicate that the CO(2,0) band is sensitive to 
both metallicity and age, and so there is an age-metallicity 
degeneracy if CO(2,0) is considered alone. Assuming that 
(1) there is not a broad galaxy-to-galaxy dispersion in luminosity-weighted age (see 
below), and (2) the luminosity-weighted age is $\sim 2$ Gyr, then the depth of this 
feature in the galaxy spectra is consistent with them having a slightly sub-solar 
metallicity similar to that of the reference spectrum. To the extent that the age-related 
assumptions hold, the agreement between the CO(2,0) depths suggests that the galaxy 
metallicities agree to within $\pm$ a few tenths of a dex, which is comparable to 
the scatter in central metallicities that can be deduced from the Mg$_2$ and CaT 
indices (Davidge 2018a). Fraser-McKelvie et al. (2018) found that the majority of the low 
mass passive galaxies in their sample -- all of which are in Virgo -- have super-solar 
central metallicities. The F2 and GMOS spectra suggest that this not the case for the 
early-type disk galaxies studied here. The GMOS spectra are of particular significance 
for metallicity determinations, as they sample H$\beta$ and H$\alpha$, thereby easing 
any age-metallicity degeneracy.

	Noise due to telluric water absorption in the 
wavelength interval that includes the $1.76\mu$m C$_2$ band is clearly evident in the 
galaxy spectra. The noise in the rest frame $1.76 - 1.79\mu$m interval 
is smallest in the NGC 4491 Region 3 and NGC 4497 Center spectra, and 
there is $\pm 1 - 2\%$ agreement with the model spectrum at these wavelengths.
While noise complicates efforts to make comparisons between the spectra of the 
other galaxies as well as with the models near $1.8\mu$m, there is a general 
tendency for the F2 spectra to agree with the reference model spectrum. 
To demonstrate this, the median of the continuum-corrected and normalized center spectra of 
all six galaxies was found, and the result is shown as a dotted line in Figures 5 and 6. 
There is a clear depression in the median spectrum that coincides with the expected 
location of the C$_2$ band, and there is good agreement between the median galaxy and the 
reference model spectrum. The mean spectrum of the center of all six galaxies shows 
similar agreement with the reference model. 

	The center of NGC 4620 has a blue $g-r$ color (Table 1), and its spectrum 
is free of line emission. Given that NGC 4620 also has the deepest H$\beta$ index among the 
six galaxies and a luminosity-weighted age of 1.5 Gyr (Davidge 2018a) then it 
might be expected to have a deeper than average C$_2$ band. It is thus worth noting that the 
C$_2$ band in NGC 4620 is not obviously deeper than in the other galaxies, although there 
is considerable noise. Given the absence of line emission, it is unlikely that 
the depth of C$_2$ in NGC 4620 has been diluted by nebular continuum emission.

	To the extent that the models track the contribution made by cool C stars 
to integrated NIR light then the comparisons in Figures 5 and 6 suggest that it is 
unlikely that the NIR light in any of these galaxies is dominated by a component 
with an age younger than 1 Gyr, otherwise a much deeper C$_2$ band 
would be seen. This is consistent with luminosity-weighted ages 
at visible/red wavelengths based on the depths of the 
H$\beta$ and H$\alpha$ lines (Davidge 2018a). The depths of the C$_2$ band in these spectra 
also do not point to a large galaxy-to-galaxy dispersion in the population
of cool C stars, again with the caveat that the models are assumed to track faithfully 
the C star content. 

	There is mixed agreement between the reference model and the galaxy spectra 
at wavelengths that cover deep atomic features. The depth of NaI$2.21\mu$m in the spectra of 
NGC 4305 and NGC 4306 matches that in the reference model. In contrast, NaI$2.21\mu$m 
in the NGC 4491, NGC 4584, and NGC 4620 spectra is markedly deeper than in the model, 
even though the comparisons discussed in Section 5.1 indicate that NaI$2.21\mu$m is 
not sensitive to metallicity and age. While there is apparent sub-structure in the NaI 
features in NGC 4306 and NGC 4491, this sub-structure is illusory and is an artifact of noise. 
Indeed, individual NaI lines would only be resolved if the assumed spectral resolution 
of these observations was markedly in error, and this is unlikely given 
that the models reproduce the width of the CO(2,0) feature in the 
galaxy spectra. Spectra of selected stars from the Rayner et al. 
(2009) library that have different NaI depths are compared in Figure 7. Spectra 
with the native resolution of the Rayner et al. (2009) sample and with that of the F2 
observations are shown. The NaI feature is smoothed considerably at the resolution of 
the F2 spectra, and it is clear that (1) the lines that make up the doublet are 
not resolved at R=400, and (2) the depth of the NaI feature with R=400 tracks the general 
behaviour seen in the R=2000 spectra. The comparisons in Figure 7 also demonstrate 
the sensitivity of NaI$2.21\mu$m to the presence of low mass stars at the spectral 
resolution of the F2 spectra.

\begin{figure}
\figurenum{7}
\epsscale{1.0}
\plotone{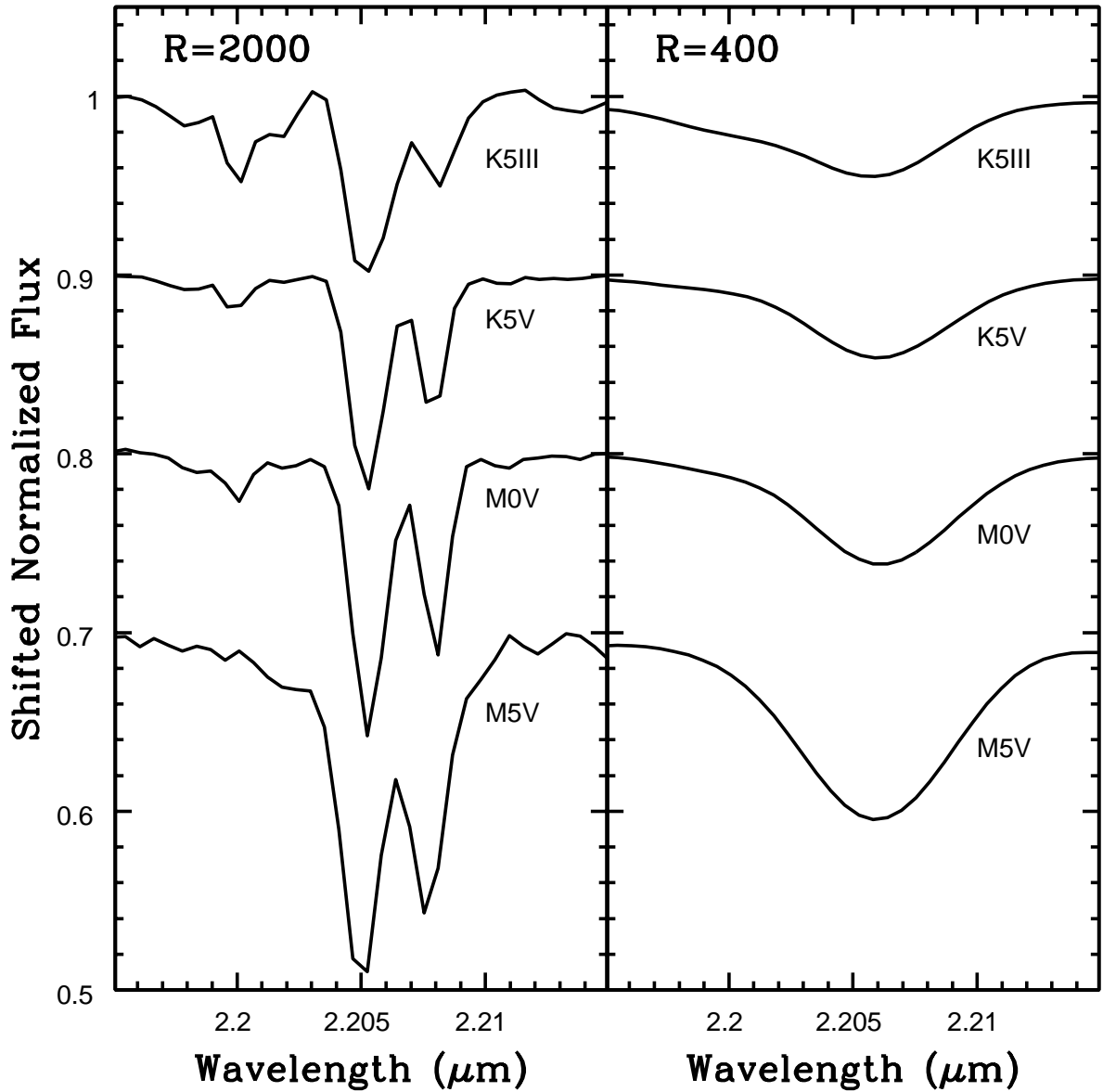}
\caption{Spectra of four stars from the Rayner et al. (2009) database at wavelengths near 
the NaI$2.21\mu$m doublet. The spectra shown are those of HD181596 (K5III), HD36003 (K5V), 
HD19305 (M0V), and Gl866ABC (M5V). The spectra in the left hand column have 
the native spectral resolution of the Rayner et al. (2009) observations, while those in 
the right hand column have the same spectral resolution as the F2 observations. 
It is evident that the lines that make up the NaI$2.21\mu$m doublet are not resolved with 
the F2 spectra, although the those spectra still track the relative depth of this feature, 
albeit with reduced sensitivity. The utility of this feature as a probe 
of low mass stars is also evident.}
\end{figure}

	In contrast to NaI $2.21\mu$m, there is general galaxy-to-galaxy 
agreement in the depth of the CaI triplet near $2.26\mu$m, although 
the NGC 4306 and NGC 4584 spectra are noisey at these wavelengths, qualifying 
conclusions about the strength of this feature in those 
spectra. Still, there is a tendency for the CaI feature to be deeper 
than predicted by the models. As with the NaI doublet, the CaI triplet is the dominant 
absorption feature at wavelengths near $2.26\mu$m, making it unlikely that 
this feature is skewed by contamination from another element. Davidge (2018a) 
found that metallicities estimated from the CaT index -- which gauges the depth 
of the Ca triplet near $0.86\mu$m -- tend to be lower than those estimated from Mg$_2$, 
indicating that the Ca triplet lines near $0.86\mu$m are weaker than in the models. 
The behaviour of the CaT index is thus contrary to what might be expected 
based on the stronger than predicted CaI$2.26\mu$m feature in the F2 data.

\section{RADIAL VARIATIONS IN STELLAR CONTENT}

	 Davidge (2018a) found radial changes in the depths of features in the 
visible/red spectra of their sample of Virgo disk galaxies, indicating that the stellar 
content is not uniformly mixed throughout these galaxies. This is not unexpected, as 
population gradients are common in galaxy disks (e.g. Gonzalez Delgado et al. 2017). 
In addition to gradients within disks, different structural components in each galaxy are 
sampled along the F2 slit, and these likely have SFHs that are distinct from each other. 
Differences between the NIR spectra of (1) the center 
region and its immediate surroundings (i.e. Region 2), and (2) the area near 
the galaxy center and the outermost areas are examined in this section.

	McDonald et al. (2009) measure the effective radii of the bulges in 
these galaxies, and find that they range from 3 to 9 arcsec in the $H-$band. 
Therefore, the comparison of light from the Center and Region 2 
notionally examines differences within the bulge, 
and this comparison is conducted for all six galaxies. 
Spectroscopic gradients might be expected if there is 
a population gradient within the bulge, or if there is a centrally concentrated 
young component and/or a nuclear star cluster. 
Many disk galaxies contain nuclear star clusters, and these have stellar contents 
that may differ from their surroundings (e.g. Carollo et al. 1998; Boker 
et al. 2004). Nuclear clusters have globular cluster-like masses, and characteristic 
sizes of 2 -- 5 parsecs, which corresponds to a few hundredths of an arcsec at 
the distance of the Virgo cluster. The closest example of a nuclear star cluster 
is the collection of stars around SgrA* (e.g. Schodel et al. 2014). It has not yet 
been determined if the Virgo galaxies studied here have nuclear clusters, although 
the central star-forming area in NGC 4491 may be an extreme example. While nuclear 
clusters in Virgo galaxies are not spatially resolved with the F2 observations, 
young or intermediate age nuclear clusters may still contribute 
significantly to the light in the central arcsec of these galaxies.

	The second set of comparisons examines differences between the central 
regions of the galaxies and Regions 3 -- 5. Regions 
3 -- 5 sample areas outside of the effective radii 
of the bulge, and so the disk contributes significantly to the integrated light. 
S/N considerations limit these comparisons to NGC 4491 and NGC 4497. 

\subsection{Comparing the Center and Region 2 Spectra}

	The Center and Region 2 spectra of all six galaxies have comparable 
S/N ratios, and the differences between the normalized spectra of these regions are 
examined in Figures 8, 9, and 10. The differenced spectra are in the sense 
Center -- Region 2. The median of the continuum-corrected and normalized 
central spectra of the six galaxies is also shown 
to aid in the identification of features. Taking the median of the spectra 
reduces random noise while also suppressing residuals from telluric features, as 
these galaxies have a wide range of radial velocities. The mean of the spectra is 
similar to the median, and the difference between the median and mean of the six spectra 
has a dispersion of $\pm 1\%$, which is comparable to the random noise levels 
in Table 2. This suggests that the median is not dominated by the spectra of the two galaxies 
with the highest S/N ratio: NGC 4491 and NGC 4497.

	The noise in the differenced spectra in Figure 8 is typically 
$< \pm 1\%$, with the differeneced spectrum of NGC 4497 having the smallest scatter. 
This noise level confirms that these spectra should be capable of detecting 
differences in the depths of features of the size discussed in Section 5 that may result 
from variations in age and/or metallicity. The low frequency wave in the differenced 
spectrum of NGC 4305 is due to uncertainties at the few percent 
level in the pseudo-continuum fit. These variations in the continuum 
occur over wavelength scales that are much wider than those expected from individual 
atomic and molecular species. 

\begin{figure}
\figurenum{8}
\epsscale{0.8}
\plotone{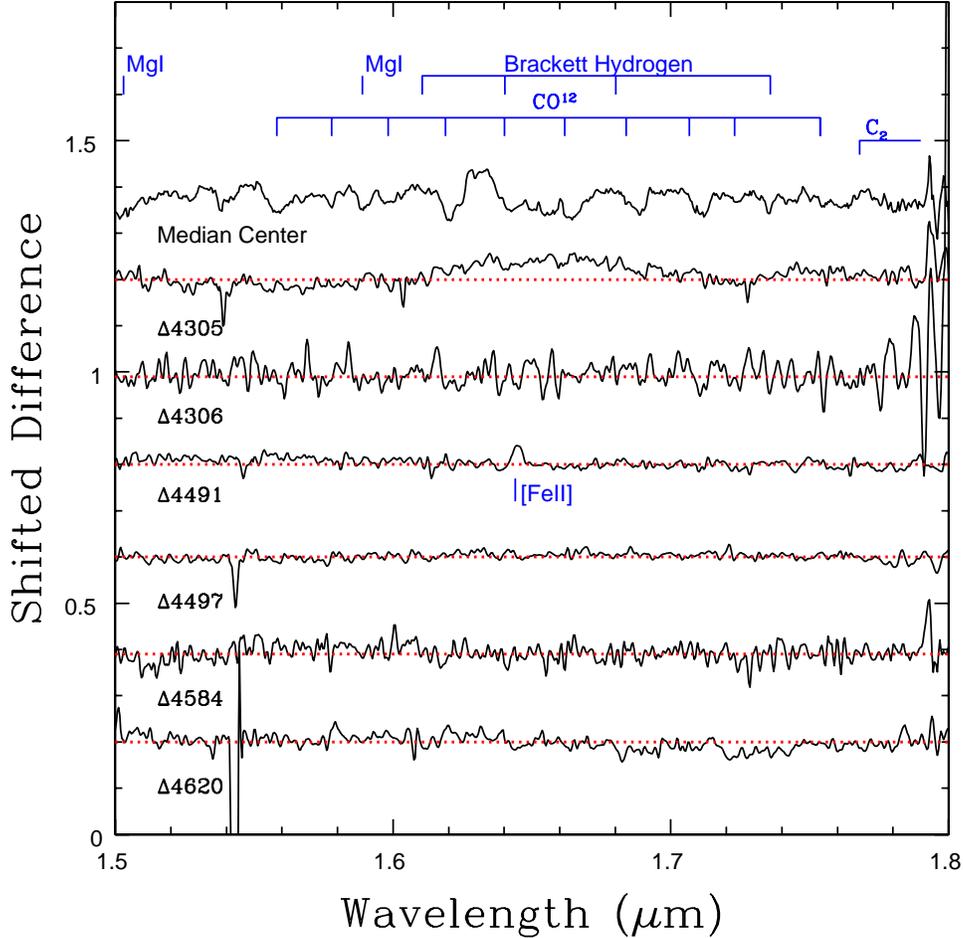}
\caption{Differences between the Center and Region 2 $H-$band spectra. 
The median of all central spectra is shown at the 
top to assist with the identification of features. The suppression of telluric 
residuals in the wavelength interval associated with C$_2$ is evident in the 
composite spectrum when compared with the spectra of individual galaxies in Figure 1. 
The differenced spectra are in the sense Center -- Region 2. 
Fractional differences can be read directly from the vertical axis 
given that continuum-corrected normalized spectra were differenced. The differenced 
spectra typically have $< \pm 1$\% scatter. The dotted horizontal red lines 
are reference sequences to assist in the detection of radial changes in absorption 
features. There is evidence for absorption with broad wavelength coverage at 
wavelengths $> 1.76\mu$m in the differenced NGC 4306 spectrum, which coincides with the 
C$_2$ band. The difference is such that C$_2$ is deeper in the Region 2 spectrum than 
in the central spectrum. There are also significant differences at the same wavelengths 
in the NGC 4497 spectrum, although in that case C$_2$ is deeper in the central 
spectrum. [FeII] emission is detected near $1.64\mu$m 
in the NGC 4491 spectrum. Brackett series emission lines are 
likely not present in the differenced NGC 4491 spectrum due to the inherent weakness of 
these features, coupled with line emission in Region 2.}
\end{figure}

	Systematic residuals that coincide with the location of the Ballick-Ramsey 
C$_2$ band are evident in the differenced spectra of NGC 4306 and NGC 4497. 
These residuals indicate that the absorption at these wavelengths 
is deeper in the center of NGC 4497 than in Region 2, while the opposite is true for 
NGC 4306. Recalling the age sensitivity of the C$_2$ feature demonstrated in 
Figure 4, the behaviour of the C$_2$ band in NGC 4497 is consistent with what might 
be expected from the GMOS spectra of that galaxy. In particular, Davidge (2018a) found 
evidence for radial age gradients in NGC 4497, in the sense of older luminosity-weighted 
ages at larger radii. Figures 16 and 17 of Davidge (2018a) indicate that the 
luminosity-weighted age of the center of NGC 4497 is 2 -- 3 Gyr. 
The contribution made by C stars to integrated NIR light at sub-solar metallicities 
peaks in populations with similar ages (Maraston 2005), and so C stars 
might be expected near the center of NGC 4497, with a diminishing C star 
frequency at progressively larger radii as mean age increases.

	As for NGC 4306, the detection of deeper C$_2$ absorption 
in Region 2 when compared with the center is perhaps 
surprising. The comparisons in Figures 3 and 4 suggest that 
if Region 2 has a higher C star density than the center then it should also have a 
different age or metallicity. However, the H$\beta$ and H$\alpha$ indices 
do not change significantly near the center of this galaxy (Figure 11 of Davidge 
2018a), while the metallicity indices in this same part of NGC 4306 are also 
more-or-less constant. There is then no supporting evidence for differences in 
the luminosity-weighted age and metallicity 
between the Center and Region 2 of NGC 4306. Differences in angular 
resolution between the GMOS and F2 datasets due to seeing are one possible cause of 
this discrepancy.

	A single emission line due to [FeII] $1.644\mu$m is seen in the differenced 
NGC 4491 spectrum. The n=12 Brackett transition has a wavelength that is 
similar to that of [FeII], and falls within the same spectral resolution element as 
[FeII] in these spectra. However, residuals from other Brackett lines 
are not seen in Figure 8, leading us to conclude that the line 
in question is [FeII], and not the n=12 Brackett transition. The [FeII] $1.644\mu$m 
emission line is a signature of SNe activity (e.g. 
Greenhouse et al. 1991). The presence of the [FeII] line in the 
differenced spectrum of NGC 4491 in Figure 8 indicates that [FeII] emission is 
concentrated near the center of this galaxy, even though hydrogen line emission 
extends into Region 2 (see below).

	The radial behaviour of the second 
overtone CO bands is explored in Figure 9, where the 
differences between the Center and Region 2 $\Sigma$CO features are 
shown. There is a dispersion of no more than $\sim 1\%$ in the differenced 
$\Sigma$CO spectrum of each galaxy. The largest dispersion is seen in the 
NGC 4584 data, although this is also the spectrum that has the lowest S/N ratio. 
Differences in the depth of the second overtone CO bands between the Center and 
Region 2 are thus modest.

\begin{figure}
\figurenum{9}
\epsscale{0.9}
\plotone{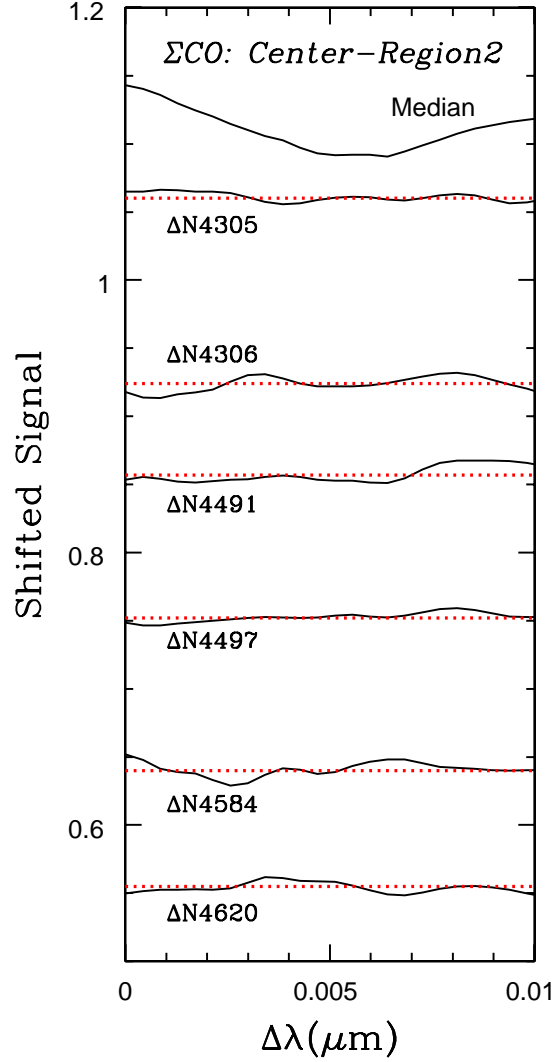}
\caption{Comparing the $\Sigma$CO second overtone features in the Center and Region 
2 spectra. $\Sigma$CO is the result of averaging 
$0.01\mu$m wide wavelength intervals centered on the (3,0), (6,3), 
(7,4), and (10,7) transitions. Differences between the 
$\Sigma$CO features in the Center and Region 2 are shown, where the difference is 
between continuum-corrected normalized spectra. The $\Sigma$CO feature in 
the median spectrum of all six galaxies is shown at the top of the panel, and the 
dotted horizontal red lines are baselines for judging structure 
in the differenced spectra. As the differences are computed from normalized spectra 
then fractional differences can be read directly from the vertical axis. 
$\Sigma$CO typically changes by $\leq \pm 1\%$ between the Center and Region 2 in these 
galaxies.}
\end{figure}

	The differenced $K-$band Center and Region 2 spectra of 
each galaxy are shown in Figure 10. With the possible exception of NGC 4497, 
the noise in these differenced spectra has a larger amplitude than in 
Figure 8, reflecting the lower S/N ratio of the $K-$band spectra. 
The model spectra discussed in Section 5 indicate that the CO(2,0) band is sensitive to 
metallicity, and CO(2,0) is weaker in Region 2 than in the center of NGC 4305 and NGC 4620. 
This is not seen in the differenced $\Sigma$CO spectra of these galaxies in Figure 9. 
There are no obvious gradients in the depths of CO(2,0) in NGC 4306, NGC 4491, NGC 4497, 
and NGC 4584. The absence of CO(2,0) gradients in the presence of age gradients is not 
unexpected unless populations with ages $\leq 1$ Gyr are involved (Figure 4). 
Finally, there are no obvious radial differences in NaI$2.21\mu$m and CaI$2.26\mu$m. 
This is not surprising given that these features are not sensitive to variations in 
age and metallicity (Section 5).

\begin{figure}
\figurenum{10}
\epsscale{1.0}
\plotone{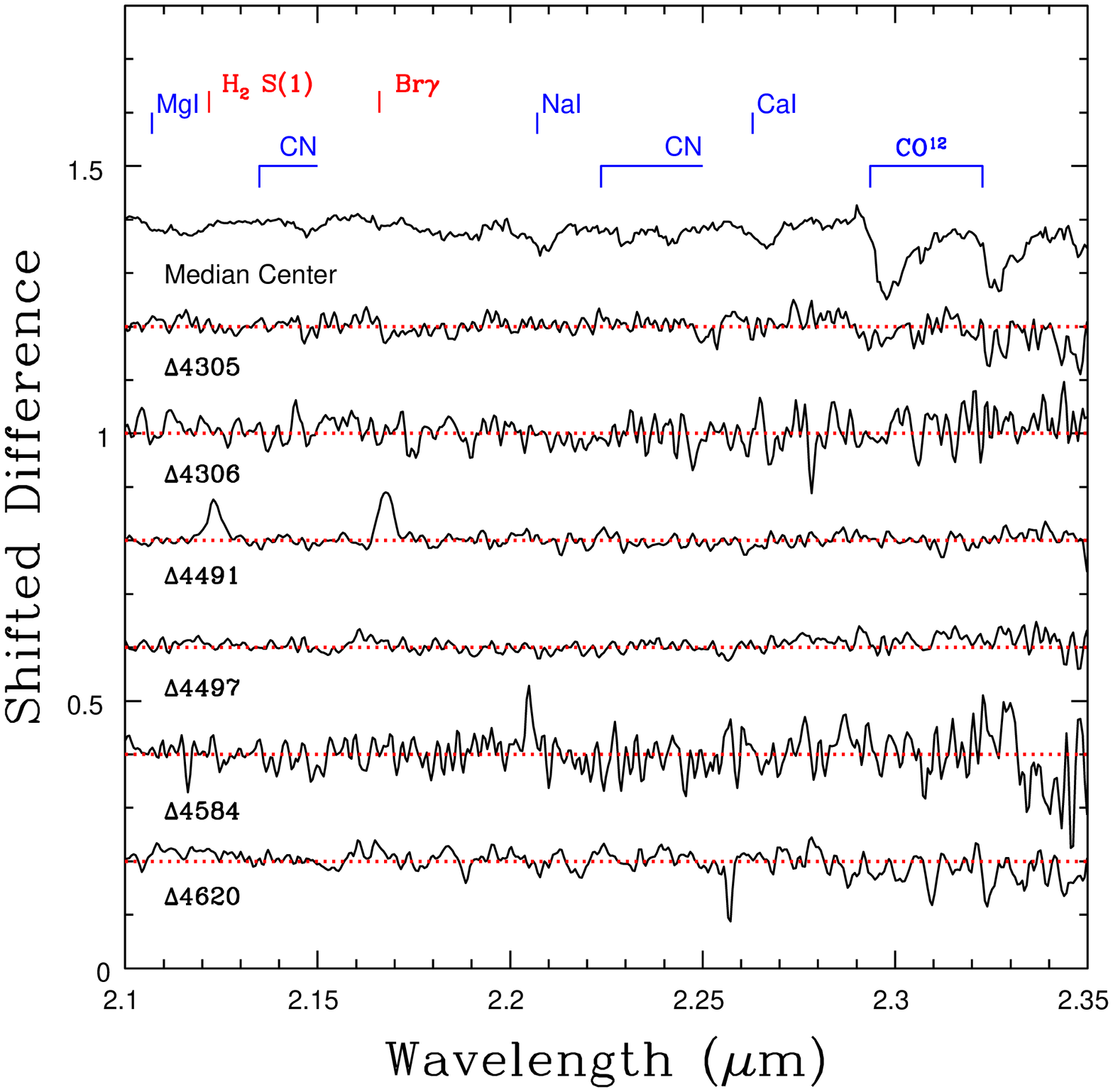}
\caption{The same as Figure 8, but showing differenced $K-$band spectra. The scatter is 
slightly larger than in Figure 8, reflecting the lower S/N ratio at these wavelengths. 
The depths of the CO bands do not change in most galaxies, although in 
NGC 4305 and NGC 4620 the CO(2,0) band is weaker in Region 2 than in the center. 
H$_2$ S(1) and Br$\gamma$ emission lines are seen in the 
differenced spectrum of NGC 4491. That these lines are weaker 
than in Figure 2 is likely because line emission is present in the Region 2 
spectrum, partially canceling signal in the central spectrum when the 
difference is taken. Extended emission of this nature may also cause the 
absence of residual Brackett emission lines in the 
$H-$band spectrum of this galaxy in Figure 8. There are 
no significant residuals associated with the first overtone CO bands in the NGC 
4491 spectrum, although CaI$2.26\mu$m in Region 2 may be weaker than in the center.}
\end{figure}

	Br$\gamma$ and H$_2$ S(1) emission lines are seen in the differenced NGC 4491 
spectrum in Figure 10, and these highlight the presence of a centrally-concentrated 
emission line region in that galaxy. There is also evidence that the line-emitting region 
extends beyond the Center region. In particular, the emission lines in Figure 
10 are weaker than in Figure 2, as would happen if emission in Region 2 nulls 
out part of the signal from the emission lines when the Region 2 and Center spectra 
are subtracted.

	The radial behaviour of the CO(2,0) feature in NGC 4491 is of interest 
given the evidence for a large centrally-concentrated young component in that galaxy. 
Despite differences in the emission line spectrum, 
the depths of the first overtone CO bands do not 
change between the Center and Region 2 in NGC 4491. This is consistent with 
nebular continuum emission not playing a major role in defining 
the depth of the CO features near the center of NGC 4491.
Moreover, if the NIR light in central regions of NGC 4491 was dominated 
by a young component that contained a large number of red supergiants and/or a large 
population of stars with an age of a few hundred Myr, which might be present if the 
galaxy has had periodic episodes of centrally-concentrated star formation 
during recent epochs, then the CO(2,0) feature should be deeper than in Region 2. That 
this is not the case suggests that the NIR light near the center of NGC 4491 is 
dominated by two idealized components: (1) a young component, with an age $\leq 10$ Myr, 
such that luminous red supergiants have not yet formed, and (2) a population with an 
age like that in Region 2. Of course, additional stellar populations 
are almost certainly present, but with the constraint that they contribute less 
to the NIR light than the two components described above.

	Emission lines are not present in the differenced NGC 4584 spectrum in 
Figure 10, confirming the observation made in Section 4 that line 
emission is weak or even abscent in the central NIR spectrum of that galaxy. 
Emission lines are present in the optical spectrum of NGC 4584, and 
images from the WISE All-Sky survey indicate that there is 
thermal emission near the center of that galaxy. The 
weakness or absence of emission lines in the NIR spectrum of NGC 4584 suggest 
that the emission in this galaxy is not powered by star formation, but could instead 
be due to LINER or AGN activity (e.g. Larkin et al. 1998; Lamperti et al. 2017). 
Kinematic measurements with sub-arcsecond angular resolution would be of interest to 
determine if there is evidence for a massive black hole at the center of NGC 4584.

\subsection{Comparing Bulge and Disk Spectra}

	In Section 5 it was shown that changes in age and metallicity 
in old and intermediate age populations are expected to have only a modest 
impact on the depths of most absorption features in the $1.5 - 2.4\mu$m interval, 
making it difficult to detect changes in the strengths of features 
if the S/N ratio is $\leq 50 - 100$. The spectra in Regions 3, 4, and 5 have lower 
S/N ratios than those in Region 2 and the Center, and low S/N ratios confound efforts to 
examine the spectra of NGC 4305, NGC 4306, NGC 4584, and NGC 4620 outside of Region 2.
However, mean spectra of NGC 4491 and NGC 4497 with a high S/N ratio can be constructed 
for Regions 3 -- 5, and the difference between the 
central spectrum of each galaxy and the mean spectrum of Regions 3 -- 5 
is shown in Figures 11 and 12. The differences are in the sense Center -- 
outer region.

\begin{figure}
\figurenum{11}
\epsscale{0.9}
\plotone{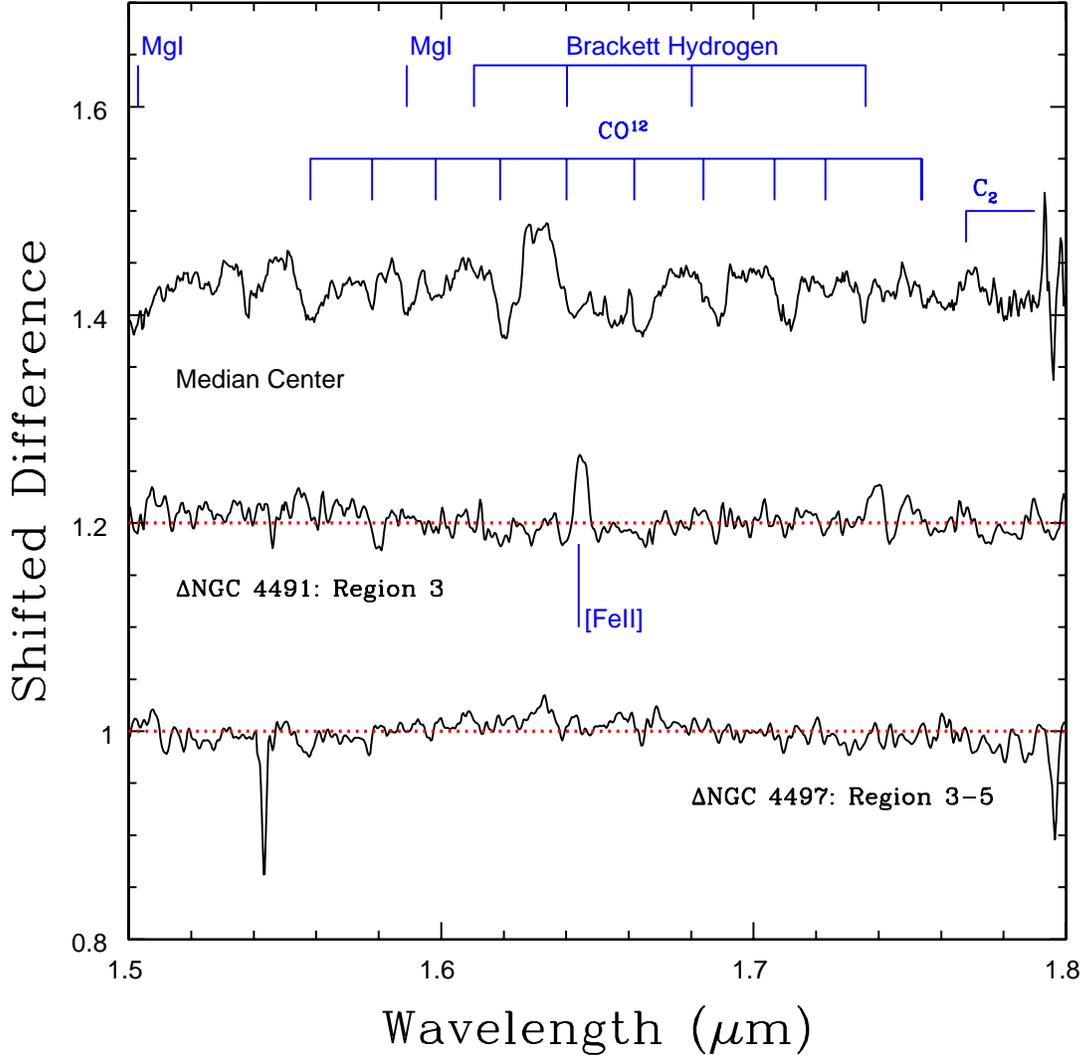}
\caption{Radial variations between the $H-$band spectra of the central and outer regions 
of NGC 4491 and NGC 4497. The differences between the central spectrum and the mean of 
the Region 3 -- 5 spectra are shown, as is the median central spectrum from Figure 8. 
The differenced spectra are in the sense center -- outer regions. The 
differenced spectra were constructed from continuum-corrected normalized spectra, and so 
fractional differences can be measured directly from the scale along the vertical axis. 
The dotted horizontal red lines provide a benchmark to assess differences in spectroscopic 
features. Noise due to telluric water absorption notwithstanding, there is a tendency for 
the differenced spectrum of NGC 4497 to drop at wavelengths that coincide with the C$_2$ 
band, signalling a tendency for centrally concentrated C$_2$ absorption in 
that galaxy. The differenced NGC 4491 spectrum also contains [FeII] $1.644\mu$m 
emission, and indicates that high-order Brackett emission lines are weak near 
the center of that galaxy.}
\end{figure}

\begin{figure}
\figurenum{12}
\epsscale{1.0}
\plotone{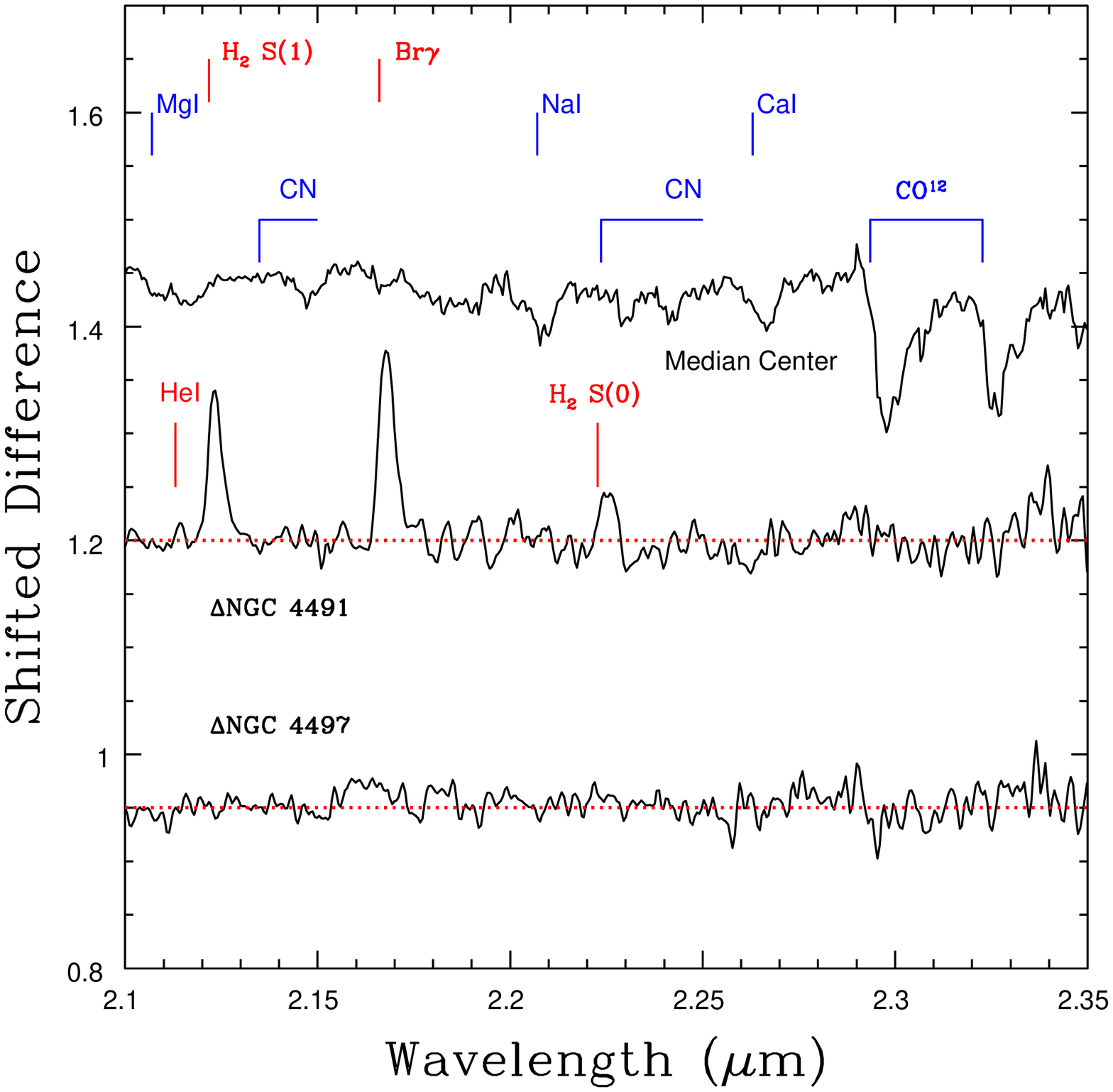}
\caption{Same as Figure 11, but showing $K-$band spectra. The locations of HeI and 
H$_2$ S(0) emission lines are indicated, and emission features are seen in the 
differenced NGC 4491 spectrum close to the expected locations of these 
features. Still, the blip that corresponds to HeI has an amplitude 
that is comparable to the noise in the spectrum, bringing 
the significance of this detection into question.}
\end{figure}

	The low amplitude of the scatter in the 
differenced spectra in Figures 11 and 12 is testament to the relatively 
high S/N ratio of the spectra of both galaxies outside of the central few arcsec. 
The constancy of the CO(2,0) band in NGC 4491 between the Center and Region 2 
also suggests that veiling from nebular continuum emission likely does not affect the 
depths of absorption features in the spectrum of that galaxy. Thus, it is likely 
that the absorption features in the center spectrum of NGC 4491 faithfully reflect the 
underlying stellar content.

	Davidge (2018a) found evidence for age gradients in NGC 4491 and NGC 4497, 
in the sense of older luminosity-weighted ages at larger radii. 
There is evidence for a possible weakening of the C$_2$ band at $1.76\mu$m 
in the Region 3 -- 5 spectrum of NGC 4497 when compared with smaller radii, 
in the form of a depression at the few percent level in the differenced spectrum. 
The models shown in Figure 3 indicate that the observed difference is larger than what 
would be expected due to a metallicity gradient. The observed trend in C$_2$ 
is consistent with Regions 3 -- 5 in NGC 4497 containing a larger fraction 
of stars with ages in excess of 2 -- 3 Gyr than near its center. 

	The differenced spectra of NGC 4491 in Figures 11 and 12
accentuate the line emission near the center of that galaxy. 
The emission lines in the differenced spectra of NGC 4491 in Figures 11 and 12
are more pronounced than in Figures 8 and 10, as expected if 
line emission is present in Region 2 (i.e. within $\pm 2$ arcsec of the 
galaxy center), but is weaker or abscent at larger radii.
The emission line strengths in Figures 11 and 12 thus can be used to 
probe the nature of the line emission near the center of NGC 4491.
There is no evidence for line emission in the differenced spectrum of NGC 4497.

	The relative strengths of [FeII], H$_2$ S(1), and Br$\gamma$ in NGC 4491 are 
not consistent with shock excitation (e.g. line strengths discussed by Sugai et al. 1997). 
This being said, the agreement with photoionization models is also not ideal. 
Black and van Dishoeck (1987) model the strengths of H$_2$ lines that are powered by 
flouresence. Considering the wavelength range sampled by F2, 
these models predict that the (1,0) transition of the H$_2$ S(1) sequence 
near $2.12\mu$m should be the strongest H$_2$ line, followed by the (1,0) S(0) 
transition. Both lines are detected in NGC 4491, and 
their relative strengths are more-or-less consistent with model predictions. 
Still, these same models also predict that the (2,1) S(1) line at 
$2.247\mu$m should have an amplitude that is just lower than that of the (1,0) S(0) 
line, and this feature is not present in the differenced NGC 4491 spectrum in Figure 12.

\section{DISCUSSION \& CONCLUSIONS}

	Long-slit NIR spectra of six early-type disk galaxies in the Virgo cluster 
that were recorded with the F2 imaging spectrograph on GS have been presented in this 
paper. These galaxies were the subject of a previous spectroscopic study at visible 
wavelengths by Davidge (2018a). The galaxies have $K-$band luminosities that are 
comparable to those of nearby late-type spiral galaxies like M33 and NGC 2403. 
Lisker et al. (2006a) assign these objects to their dEdi class, and 
identify five as dwarf S0s. While the galaxies appear to be largely devoid of 
cold interstellar material, NGC 4491 and NGC 4584 have emission at $24\mu$m 
that suggests there are significant amounts of hot dust in their central regions. 

	The spectral resolution of the F2 data is sufficient to allow features 
such as the C$_2$ band at $1.76\mu$m, the CO (2,0) band head, the NaI$2.21\mu$m 
doublet, and the CaI$2.26\mu$m triplet to be detected, and these are used to 
investigate the stellar contents of these systems. The goal is not to derive independent 
luminosity-weighted ages and metallicities, but rather to check for consistency with 
properties found from the visible/red spectra. In addition, as the NIR spectra of faint 
objects can contain systematic residuals introduced by telluric features that 
can skew index measurements, the radial properties 
of the spectrum of each galaxy are examined in a 
differential manner in an effort to reduce any such systematic effects.

	The study of NIR spectra is complementary to work 
at visible wavelengths as the mix of stellar types that dominates the NIR light 
differs from that at shorter wavelengths. A large 
fraction of the NIR light from systems with intermediate and old ages originates 
from luminous highly evolved stars, whereas at visible wavelengths 
there are significant contributions to the light from 
objects that span a range of evolutionary states, including the main sequence turn-off and 
giant branch (e.g. Figure 13 of Maraston 2005). Spectra at NIR wavelengths are also less 
affected by dust extinction. NIR spectra thus provide additional 
insights into the stellar contents and past histories of these galaxies. 

	The sensitivity of features in the NIR 
to changes in age and metallicity have been assessed using models from the EMILES 
compilation. These models indicate that changes in age and metallicity that are consistent 
with what might be expected from the examination of spectra at visible wavelengths alter the 
depths of features in the NIR by a few percent. NIR spectra with a S/N ratio of 
50 -- 100 have thus been extracted within a few arcsec of the centers 
of all six galaxies, while in NGC 4491 and NGC 4497 it has been possible to examine 
their spectra out to angular offsets that extend to roughly the half light radius.

	The main conclusions of this paper are as follows:

\vspace{0.3cm}
\noindent{1)} The F2 spectra reveal galaxy-to-galaxy differences in stellar content that 
appear not to be related to age or overall metallicity. Comparisons with E-MILES 
models that use the BaSTI isochrones are described in Section 5, and 
the NaI$2.21\mu$m feature in these models is not sensitive to changes in 
age and overall metallicity. As the NaI$2.21\mu$m lines are not resonance transitions 
then this feature is likely not affected by absorption from the ISM. 
It is thus significant that the NaI$2.21\mu$m feature in the spectra of NGC 4491, NGC 
4584, and NGC 4620 is (1) deeper than in the other three galaxies, and (2) 
deeper than predicted by models that assume a solar chemical mixture. 

	The differences in the depth of NaI$2.21\mu$m are not 
restricted to the centers of these galaxies. The comparisons 
between the Center and Region 2 spectra discussed in Section 6
suggests that the depth of Na$2.21\mu$m does not vary within the central 
few arcsec of these galaxies. The Region 3 spectrum of NGC 4491 was used to 
assess the strength of NaI$2.21\mu$m in that galaxy, further confirming that 
the enhanced NaI$2.21\mu$m feature is not restricted to the galaxy centers. 

	A property of NGC 4491, NGC 4584, and NGC 4620 that is 
not shared with the other galaxies is that all three have similar central
$g-i$ colors (Table 1). While emission lines are present near the centers of 
NGC 4491 and NGC 4584, this is not the case for NGC 4620, although the Balmer 
line depths in the latter are indicative of a young age. 
The Sersic indices of the central regions of NGC 4584 and NGC 4620 are consistent with those 
of classical Sa and Sb galaxies (McDonald et al. 2011), 
while the other four galaxies have central Sersic indices that are appropriate for 
late-type systems. The interpretation of Sersic 
profiles in galaxies with a centrally-concentrated light profile due to a 
star-forming nucleus or an AGN is not clear, although 
a centrally-peaked light distribution will bias Sersic indices to higher values 
(i.e. in the sense of making them more consistent with those of classical bulges).

	The discovery of galaxies with deep NaI$2.21\mu$m absorption is not new. 
This feature has been found to have a range of depths among galaxies, tending 
to be deeper than expected in many early-type galaxies if a solar chemical mixture and 
a solar neighborhood mass function are assumed. Rock et al. (2017) examine the strength 
of NaI$2.21\mu$m in early-type galaxies, and investigate models that have a 
non-solar [Na/Fe] and a bottom-heavy mass function. They conclude that the deep 
Na $2.21\mu$m lines seen in their sample are due to [Na/Fe] $> 0$ 
coupled with a possible bottom-heavy mass function. They suggest that a 
non-solar [C/Fe] may also contribute to deep NaI$2.21\mu$m absorption. 

	While NaI lines are sensitive to surface gravity, in the sense of 
being deepest in stars with the highest surface gravities at a 
fixed effective temperature, some NaI transitions are more sensitive to surface 
gravity than others, and this information can be used to constrain the source of the 
bi-modal NaI$2.21\mu$m behaviour in these galaxies. Conroy \& van Dokkum (2012) discuss 
models that indicate that the sensitivities of NaI$2.21\mu$m to [Na/Fe] and the 
slope of the mass function at low masses are intermediate between those of 
NaD (sensitive to [Na/Fe]) and NaI$1.14\mu$m (sensitive to the mass function). 
If the deep NaI$2.21\mu$m features in NGC 4491, NGC 4584, and NGC 4620 are due to 
super-solar [Na/Fe] then the NaD lines in their spectra should be deeper than in 
the other galaxies. In fact, Davidge (2018a) found 
that the NaD lines in these three galaxies have equivalent widths that 
are smaller than those in the other three galaxies. 
However, NaD is not an ironclad probe of stellar content, as it can be affected by 
non-stellar sources. For example, the NaD lines are resonance 
transitions, and so are susceptible to absorption from 
interstellar material. This being said, the presence of interstellar absorption will 
strengthen the lines in NGC 4491, NGC 4584, and NGC 4620, and not make them weaker. It is 
also unlikely that interstellar absorption has deepened the NaD lines in the three 
galaxies in which NaI$2.21\mu$m matches the models, as these galaxies do not appear to 
have a significant ISM.

	Aside from the potential for a contribution from the ISM, the depth 
of NaD is also sensitive to age. The strength of this feature in models 
weakens as age decreases (e.g. Figure 12 of Conroy \& van Dokkum 2012), tracking 
the temperature of the main sequence turn-off. However, Balmer lines in the GMOS spectra 
are consistent with luminosity-weighted ages of $\sim 2$ Gyr, and the depth of NaD in 
NGC 4620 is consistent with that predicted by the models (e.g. Figure 19 of 
Davidge 2018a). The depth of the NaI features are also sensitive to chemical 
mixture (Conroy \& van Dokkum 2012), with NaI$2.21\mu$m being sensitive to [C/Fe] 
(Rock et al. 2017). Davidge (2018a) suggested that these galaxies may have solar chemical 
mixtures, although the inconsistent metallicities obtained from the Mg$_2$ and 
CaT indices measured by Davidge (2018a) hint that this may not hold for all elements.

	The NaD lines may also be partially filled by emission from a nebular continuum. 
In fact, NGC 4491 and NGC 4584 both have a centrally-concentrated 
emission line component at visible/red wavelengths. However, 
NaD in NGC 4491 and NGC 4584 is more-or-less constant with radius 
(e.g. Figure 14 of Davidge 2018a). If NaD near the galaxy center was affected 
by a nebular continuum or the presence of young stars then NaD would deepen 
towards larger radii, and such a trend is not seen. As for NGC 4620, there are no emission 
lines in its spectrum, and NaD also does not vary with radius. NaI$2.21\mu$m 
does not vary between the Center and Region 2 in any of these galaxies, while 
NaI$2.21\mu$m in NGC 4491 is deeper than predicted by the models 
in Region 3, which is outside of the central star-forming region.

	The problems with the depth of NaD notwithstanding, any 
mechanism that involves Na enhancement to explain the deeper than expected 
NaI$2.21\mu$m lines found here must do so over a large spatial scale, 
and not just over a localized part of these galaxies.
Na is produced in high temperature conditions, such as in 
Type II supernovae (SNeII) and in massive AGB stars during hot bottom burning (HBB). 
Na can be either formed or destroyed during HBB, depending on 
the temperature at the bottom of the convective envelope (e.g. Ventura \& D'Antona 2011). 
A problem with producing large-scale Na enrichment with AGB stars is then that a 
remarkably fine-tuned mass function that favors the progenitors of only very massive AGB stars 
is required. 

	Kobayashi et al. (2006) examine the chemical enrichment of elements produced 
by massive stars. The production of Na with respect to other elements climbs with 
progenitor mass among stars with a solar metallicity, and a similar trend 
is seen for Al. Na (and Al) enrichment might then result from 
a mass function that favors very massive stars. 
Super massive stars (SMSs) are then one conjectural means of producing large-scale 
Na enrichment during early epochs. SMSs are proposed to form in deep potential 
wells with high gas accretion rates where runaway stellar collisions occur (Gieles et 
al. 2018), presumably during very early epochs. Denissenkov \& Hartwick (2014) discuss 
SMSs as a possible source of chemical abundance trends among stars in globular clusters, 
and find good agreement between observed and modelled abundance ratios. Their models 
reproduce the [Na/Fe] $vs$ [O/Fe] anticorrelation seen in globular clusters. However, 
it is not clear if such a mechanism could produce sufficient quantities of Na that 
must then be distributed over large spatial scales.

	Taken at face value, the relative strengths of the NaD lines suggest 
that the deep NaI$2.21\mu$m lines in NGC 4491, NGC 4584, and NGC 4620 are likely a 
consequence of a bottom-heavy mass function, rather than a super-solar Na abundance. 
Such a mass function is unexpected in light of 
the relation between velocity dispersion and mass function exponent that has been 
found among early-type galaxies (e.g. Rosani et al. 2018, Spiniello et al. 2014, Conroy 
et al. 2013, Cappellari et al. 2013, Ferreras et al. 2013). However, the alternative 
of proposing a Na enhancement in NGC 4491, NGC 4584, and NGC 4620 has difficulties 
given the relative depths of NaD in the six galaxies. 
Other spectroscopic features will provide additional insights into the mass 
function of these galaxies. For example, the NaI$0.82\mu$m doublet is sensitive to surface 
gravity. While the spectra discussed by Davidge (2018a) cover this wavelength region, the 
spectral resolution is too low to allow meaningful constraints to be drawn, 
and observations of these galaxies at spectral resolutions in excess of 
1000 would be of interest to measure the depth of this feature.
A study of the NaI$1.14\mu$m line in all six galaxies would also be of interest to 
further constrain the contribution made by low mass stars to the NIR light. 
Finally, the AlI$1.31\mu$m feature is also of interest, as Al has a nucleosynthesis 
pedigree that is similar to that of Na in massive stars -- if Na is enhanced 
then this might also be reflected in the depths of Al lines. 

\vspace{0.3cm}
\noindent{2)} The NIR spectra indicate that the ionizing radiation 
that causes the emission in the central regions of NGC 
4491 and NGC 4584 may originate from very different sources. The spectra of NGC 4491 and 
NGC 4584 have prominent emission lines at visible wavelengths (Davidge 2018a), and 
these galaxies have similar central $g-i$ colors (Table 1). These are also the only two 
galaxies to be detected in WISE W4 images. Despite these similarities, the 
NIR spectra of NGC 4491 and NGC 4584 are very different: whereas there 
are emission lines in the NIR spectrum of NGC 4491, 
no emission lines are detected in the NIR spectrum of NGC 4584. 

	The relative strengths of the H$_2$ (1--0) S(1) and 
Br$\gamma$ lines in NGC 4491 are consistent with those found 
in star-forming galaxies, although the non-detection of the (2,1) S(1) H$_2$ line 
is a possible problem (Section 6). As for the center of NGC 4584, the absence of NIR 
emission in that galaxy could signal an excitation mechanism that is not related to 
star formation, such as an AGN. The wavelength coverage and spectral resolution of 
the GMOS spectra discussed by Davidge (2018a) are such that 
excitation diagnostics like [OII]3727 and [NII]6563 (e.g. Baldwin, 
Phillips, \& Terlevich 1981) can not be measured from those data. Visible and red spectra 
of NGC 4584 with broader wavelength coverage and higher spectral resolution would thus 
be useful to further examine the excitation mechanism in that galaxy. 
Images and spectra of NGC 4584 that have sub-arcsec angular resolution will also provide a 
direct means of determining the angular extent of the emission. 
We note that NGC 4584 is not barred.

\vspace{0.3cm}
\noindent{3)} The NIR spectra are consistent with the center of 
NGC 4491 being an extreme star-forming environment. The $1.64\mu$m [FeII] emission 
in the NIR spectrum is an indicator of a high 
level of supernovae activity, the progenitors of which were likely massive, hot stars. 
If confirmed by spectra with a higher S/N ratio, then the presence of HeI emission near 
$2.11\mu$m would suggest that there may be a population of extremely hot stars, 
including Wolf-Rayet stars, near the center of NGC 4491. 
A large population of hot stars is consistent with 
the dust temperature of NGC 4491 estimated by Auld et al. 
(2013), which is the highest among galaxies examined in the Herschel Virgo Cluster Survey. 
The MIR SED of NGC 4491 in the NASA Extragalactic 
Database \footnote[10]{http:ned.ipac.caltech.edu} is similar to that of NGC 5253, 
a nearby dwarf galaxy that harbors vigorous on-going star formation 
(e.g. Alonso-Herrero et al. 2004).

	The evidence for concentrated central star formation notwithstanding, 
NGC 4491 is not unique among early-type galaxies in cluster environments, and we suggest 
that it may be related to the MIR enhanced galaxies (MIREGs) identified 
in the Virgo and Coma clusters by Riguccini et al. (2015). 
The SEDs and locations of MIREGs in the outer regions 
of both clusters suggest that they may be experiencing an episode of intense 
star formation after falling into the cluster environment. Of the three MIREG 
galaxies found in Virgo, the integrated $r$ brightness and $g-r$ color of 
one -- NGC 4344 -- are comparable to those of NGC 4491. The ratio of IRAS $25\mu$m to 
$K-$band flux in NGC 4491 is $2 \times$ that in NGC 4344, suggesting that NGC 4491 may 
be a more extreme MIREG than NGC 4344. This might be a consequence 
of the location of NGC 4491 within Virgo; unlike the majority of MIREGs, 
NGC 4491 has a projected location on the sky that is relatively close to M87 
(Table 1). While the projected location of an object in a 
three-dimensional structure is not a certain indicator of its physical location 
within that structure, the lower-than-average radial velocity 
of NGC 4491 in Table 1 suggests that its orbit may have been dynamically processed 
by the cluster environment, perhaps due to an interaction with a much larger system. 
Such an interaction could have triggered the central star formation seen today. 

	NGC 4491 was not considered by Riguccini et al. (2015) in their census 
of MIREGs because of its morphological classification. The sample of objects examined by 
Riguccini et al (2015) drew on that discussed by Temi et al. (2009), who 
selected galaxies of type E and S0 as assigned by Bingelli et al. (1985). As 
noted in Table 1, NGC 4491 was assigned type SBa(s) by Bingelli et al. (1985). 
However, that Lisker et al. (2006a) conclude that NGC 4491 is a 
dwarf S0 re-opens the possibility of considering it as a MIREG.

	The nature of MIREGs notwithstanding, a nucleus with an elevated 
SFR may not on its own indicate that a galaxy is a recent addition to 
Virgo. The depth of Balmer absorption lines outside of the central regions of 4491 is 
consistent with a relatively old age (e.g. Figure 16 of Davidge 2018b). 
That there is no evidence for a large intermediate age population in the disk of 
NGC 4491 suggests that any on-going star formation 
has been at a low enough level so as not to dominate the integrated 
light at visible wavelengths, although a young frosting of stars is present in many 
early-type galaxies (e.g. Trager et al. 2000; Ford \& Bregman 2013). 
This leads us to suggest that the recent star formation near the center of NGC 4491 
may be a spatially isolated event, perhaps akin to what is seen in the centers 
of nearby spiral galaxies (e.g. Davidge \& Courteau 2002; Georgiev \& Boker 2014), 
dwarf lenticulars (e.g. Seth et al. 2010; Davidge 2015a), and dwarf galaxies 
in Virgo (Cote et al. 2006). 

\vspace{0.3cm}
\noindent{4)} The stellar contents in the central regions of the galaxies are not 
tied to the presence of a bar at the present day. 
Intermediate mass disk galaxies like those studied here are susceptible to bar instabilities, 
and bars can form and buckle throughout their lifetimes (e.g. Kwak et al. 2017). 
As the current sample consists of early-type disk galaxies then there might be a 
bias towards galaxies that had a bar in the past but that has since buckled to form a 
pseudo-bulge. Indeed, despite having early-type morphologies (Table 1), NGC 4305, NGC 
4306, NGC 4491, and NGC 4497 have central Sersic profiles that are not consistent with 
those expected for classical bulges (McDonald et al. 2011).

	A connection between the central SFR and a bar might 
be expected in a galaxy that contains large quantities of cool 
gas in the disk, as a bar can channel that gas into the central regions, 
which in turn could result in central star formation. While three of the six galaxies have a 
bar (NGC 4306, NGC 4491, and NGC 4497), the central stellar contents of these objects are 
diverse. NGC 4491 shows evidence of on-going central star formation, while Davidge (2018a) 
found that NGC 4497 has the oldest luminosity-weighted age of the six galaxies. 

	While star formation might be sustained near the geometric center of a bar due 
to gas deposition, bars also play a role in suppressing star formation 
in disks. Observational evidence for a bar-related 
decrease in the disk SFR comes from the red colors of 
barred spiral galaxies (e.g. Vera et al. 2016). Fraser-McKelvie et al. (2018) find 
that passive spiral galaxies have a higher bar frequency than star-forming 
spirals, while James \& Percival (2018) find that the area swept by bars tends to be 
devoid of recent star formation. Khoperskov et al. (2018) simulate the effect of 
bars on gas disks, and find that the SFR in their models drops after bar formation. 
This drop in the SFR occurs because the bar increases turbulence in the disk, heating the 
gas and thereby hindering fragmentation and collapse. This bar-driven turbulence enlarges 
the vertical extent of the gas disk in these models, but the gas is not lost from the 
galaxy unless it is removed by some other mechanism, such as tidal interactions or ram 
pressure. 

\vspace{0.3cm}
\noindent{5)} The wavelength region near $1.76\mu$m in the galaxy rest frame covers 
the Ballick-Ramsey C$_2$ band and is reproduced by the E-MILES models that are based on 
the BaSTI isochrones with a scaled-solar chemical mixture. This agreement is 
worth noting given the uncertainties in the physics used to model the structure 
of stars in advanced stages of evolution and the contribution that they make to the 
spectra. These uncertainties affect the predicted incidence of highly 
evolved AGB stars and/or their spectroscopic characteristics in models of integrated 
light. This being said, with the possible exception of NGC 
4491, the spectra are noisey at the wavelengths in question. 
Observations made from a drier site, where H$_2$O absorption 
is less of a factor and more stable with time, should yield spectra 
at these wavelengths with lower noise levels. 

	Zibetti et al. (2013) examined the NIR spectra of a sample of post-starburst 
galaxies to search for spectroscopic signatures of C stars. The galaxies in their 
sample are at redshift 0.2, and so the $1.76\mu$m C$_2$ band 
is redshifted into the $K-$band, where it is less affected by deep telluric 
H$_2$O absorption. Zibetti et al. (2013) do not detect deep C$_2$ features, 
even though these galaxies were selected to be sites of recent 
large-scale star formation. However, the S/N ratio of their spectra are such that 
absorption near rest frame $1.76\mu$m with a depth like that seen in the present 
sample of Virgo galaxies would likely not be detected.

\vspace{0.3cm}
\noindent{6)} The depths of the first overtone CO bands are consistent with 
the galaxies having similar luminosity-weighted metallicities. The models discussed in 
Section 5 indicate that the depths of the CO bands are subject to an age-metallicity 
degeneracy. However, this is a concern only for systems with ages that are ruled out 
in the current sample of galaxies based on the depth of Balmer lines at visible and 
red wavelengths; hence, the main driver of the depth of the first 
overtone CO bands in these galaxies is probably metallicity. 
The uniform depths of the CO(2,0) band head is more-or-less consistent with the modest 
scatter in [Fe/H] found by Davidge (2018a) in the central regions of these galaxies. We do 
not find evidence for super-solar luminosity-weighted central metallicities. 

\vspace{0.3cm}
\noindent{7)} There is galaxy-to-galaxy agreement in the depth of 
the CaI$2.26\mu$m triplet, which is uniformly deeper 
than predicted by the reference model. The models discussed in Section 5 indicate that 
the depth of CaI$2.26\mu$m is slightly sensitive to variations in age, in the sense of 
becoming deeper for ages $< 2$ Gyr, and this is likely a surface gravity effect. However, 
an age difference with respect to the reference model is likely not an explanation 
for the depth of CaI$2.26\mu$m in the galaxy spectra, as the depths of Balmer lines at visible 
wavelengths and the depth of the C$_2$ band point to luminosity-weighted ages 
that more-or-less agree with that of the reference model. Ca is an $\alpha$ element, and the 
difference in the depth of the CaI$2.26\mu$m feature with respect to the reference 
model might instead point to a non-solar chemical mixture in these galaxies.

	We close the paper by re-visiting the evolutionary status of the galaxies 
in light of information gleaned from the F2 spectra. Davidge (2018a) 
suggested that these galaxies likely formed as late-type spirals 
in a low density environment, and that they were stripped of gas after falling 
into the cluster. Pseudo-bulges would have presumably formed following the formation 
and collapse of bars, although the inner region light profiles of two of the 
galaxies -- NGC 4584 and NGC 4620 -- can be fit with Sersic indices 
that are suggestive of classical bulges. The entry into the cluster environment occured 
only after a stable disk had formed that had had time to experience substantial 
chemical evolution. 

	The evolutionary scheme described by Davidge (2018a) 
relies on the presence of fossil disks, that are now 
largely devoid of the recent large-scale star formation that is characteristic of a 
late-type spiral galaxy. Such disks are clearly present in the galaxies studied 
here. Another characteristic that is consistent with 
an external formation model is the chemical mixture, which Davidge (2018a) found 
to be near-solar. A solar abundance mixture indicates that a system was able to 
retain star-forming material long enough for SNe I to enrich the interstellar material 
and form a large population of stars (i.e. $\geq 1$ Gyr) -- 
gas removal did not happen suddenly after the disks formed. While the CaI$2.26\mu$m 
feature appears to be deeper than predicted by the models, thereby suggesting a 
non-solar chemical mixture, the galaxy-to-galaxy agreement in the depth of this feature is 
consistent with a common Ca abundance among the galaxies.

	The galaxies studied here have similar red $g-i$ colors and 
integrated brightnesses, although the former is partly a selection effect 
due to their early-type morphologies. The spectroscopic 
properties of these galaxies at visible wavelengths 
indicate that there is a modest dispersion in luminosity-weighted age 
and metallicity, as might be expected for galaxies evolving in a dense 
cluster that has accreted galaxies over a range of epochs. 
Still, the NIR spectra contain clues that the six galaxies may not have stellar contents 
that differ solely due to variations in luminosity-weighted age and metallicity, 
and so may not share a common pedigree. The detection of deep NaI$2.21\mu$m in three of these 
galaxies argues that their stellar contents differ from those in the majority of 
late-type disk galaxies. Based on the strengths of NaD absorption, 
a bottom-heavy mass function in NGC 4491, NGC 4584, and NGC 4620 is favored to explain 
the depths of NaI$2.21\mu$m in these galaxies. To date, bottom-heavy 
mass functions have been found in massive galaxies in 
cluster environments (e.g. Rosani et al. 2018, Spiniello et al. 2014; 
Conroy et al. 2013; Cappellari et al. 2013; Ferreras et al. 2013). However, there are 
hints that factors other than total mass may be at play, as Zieleniewski et al. 
(2017) find a range of mass functions in bright Coma cluster galaxies, 
while Meyer et al. (2019) find a solar neighborhood-like mass function 
in the lenticular/elliptical galaxy M85, which is on the periphery of the Virgo cluster 
and is more massive than the galaxies studied here.
We speculate that any diversity in NaI$2.21\mu$m characteristics among the 
six galaxies in this sample may reflect differences in their formation environment, perhaps 
related to the sub-structure within the Virgo cluster (e.g. Bohringer et al. 1994). 

	Fraser-McKelvie et al. (2018) discuss quenching mechanisms 
in passive spiral galaxies. The low mass part of their sample consists 
exclusively of disk galaxies in Virgo, and includes NGC 4305. They conclude that 
these galaxies have relatively young luminosity-weighted ages and super-solar 
metallicities. They further suggest that these galaxies entered Virgo 1 -- 2 Gyr in the 
past, and that star formation was quenched by ram pressue 
stripping or by strangulation of the gas supply. 
A cautionary note is that the Fraser-McKelvie et al. spectra cover only 
the central regions of each galaxy, and the GMOS and F2 long slit spectra indicate that the 
central properties of disk galaxies are not proxies for the rest of the galaxy.

	Gradients in luminosity-weighted age found at visible wavelengths provide clues 
into how long the galaxies might have been evolving in a cluster environment. The age 
gradients found by Davidge (2018a) are in the opposite sense to those seen in late type 
disk galaxies, where the fractional contribution of younger stars {\it increases} towards 
larger radii (e.g. Gonzalez-Delgado et al. 2017). The radial age 
trend found by Davidge (2018a) is consistent with star formation 
being quenched first at large radius and then with quenching moving progressively inwards 
in all six galaxies. Such a trend might result from the initial removal of gas 
in areas where it is least tightly bound to the galaxy - the low density outer 
regions - before gas is stripped from denser regions. The ages estimated 
by Davidge (2018a) at the largest angular offsets from the galaxy centers indicate 
that large-scale star formation ceased many Gyr in the past, although it should be recalled 
that luminosity-weighted ages have been measured here, and these 
provide only loose constraints on when large-scale star formation last occured. 
Still, the age gradients in these galaxies indicate that the time scale for 
the suppression of star formation likely spanned several Gyr. This is consistent 
with the quenching time scale for anemic disk galaxies in general (Schawinski et al. 2014).

	Ruggiero \& Lima Neto (2017) simulate the evolution 
of galaxies in cluster environments, and find that 
if there is a kinematically cool cluster core then galaxies with a Milky Way-like mass can 
be stripped of gas in a single cluster crossing time. In contrast, multiple passes are 
required to strip gas in a cluster that lacks a cool core, such as Virgo. 
Thus, if the present-day properties of the early-type disk galaxies discussed here 
are due to the removal of gas by the cluster environment then they may have 
been in the Virgo cluster for multiple crossing times. This corresponds 
to at least a few Gyr given the radial extent of the cluster and the typical velocities of 
its member galaxies (e.g. Tully \& Shaya 1984). A timescale for gas removal of a few Gyr is 
consistent with the luminosity-weighted ages estimated by Davidge (2018a) from 
Balmer lines at intermediate radii.

	The radial age trends could be driven by mechanisms other than 
the stripping and/or disruption of the gas supply due to the cluster environment. 
For example, the gas in a spiral galaxy that is on its 
first pass through a cluster will likely be compressed by the intergalactic 
medium, and large-scale star formation might ensue. A large fraction of the gas in 
such a galaxy might then be consumed by star formatio, which might be 
quenched by the removal of gas by winds and/or the formation of a bar. Such 
star-forming activity in starburst galaxies typically dies out first at large radii, 
and then shrinks in size, with the last vestages of elevated activity occuring near the 
galaxy center (e.g. Soto \& Martin 2010). Observations of low mass galaxies suggest 
that elevated levels of star formation persist for at least a few hundred Myr 
to more than a Gyr (e.g. McQuinn et al. 2010). The 
result will be an age gradient with the oldest luminosity-weighted ages at large radii, 
as is seen here. As with ram stripping, the hypothesized star burst and subsequent 
quenching of star formation would have commenced many Gyr in the past to produce 
the luminosity-weighted disk ages found here. 

	A potential problem with starbursts as a driver of gas exhaustion 
is that while starburst activity is seen in some Virgo spirals (Koopman 
\& Kenney 2004a), it is not common (e.g. Crowl \& Kenney 2008).
Moreover, if starburst activity were the cause of radial trends in these galaxies 
then a population of starbursting dEs might also be expected to be found in Virgo. 
While we are not aware of such a population, there are dEs that harbor 
blue cores that are attributed to recent star formation, much like in 
NGC 4491. These objects account for 15\% of the brightest dEs, and have morphological 
characteristics that are reminiscent of thick disks (Lisker et al. 2006b), suggesting a 
possible connection to the galaxies studied here. Evidence for large-scale star formation 
at larger radii is lacking in those objects, suggesting that they are not post-starburst 
systems. This is consistent with their nuclear star formation being triggered 
by some other process.

	Establishing the SFHs of the disks during intermediate epochs 
will provide direct constraints on when these galaxies first entered a 
cluster environment and experienced the events that subsequently shaped their evolution. A 
study of the resolved stellar contents of these systems would be of particular 
interest. Simulations discussed by Schreiber et al. (2014) indicate that 
it should be possible to resolve individual RGB and AGB stars 
in Virgo galaxies using adaptive optics systems mounted on 
the next generation of very large optical/NIR telescopes, and 
the resulting color-magnitude diagrams could be used to place constraints on their SFHs. 
A caveat is that while the identification of epochs 
with elevated SFRs is of obvious interest, such events may not always be 
due to cluster-driven evolution. For example, the SFRs in the disks of `typical' 
spiral galaxies experienced a rejuventation $\sim 4$ Gyr in the past 
(Gonzalez-Delgado et al. 2017). If the galaxies entered the 
Virgo environment since that time then evidence for an uptick in the SFR $\sim 4$ Gyr 
ago might then be a relic of their evolution prior to entering the cluster.

\acknowledgements{It is a pleasure to thank the anonymous referee for providing  
comprehensive reports that greatly improved the paper.}


\begin{references}

\reference{}Alonso-Herrero, A., Takagi, T., Baker, A. J., Rieke, G. H., Rieke, M. J., Imanishi, M., Scoville, N. Z. 2004, ApJ, 612, 222

\reference{}Baillard, A., Bertin, E., de Lapparent, V., et al. 2011, A\&A, 532, 74

\reference{}Baldwin, J. A., Phillips, M. M., \& Terlevich, R. 1981, PASP, 93, 5

\reference{}Bamford, S. P., Nichol, R. C., Baldry, I. K., et al. 2009, MNRAS, 393, 1324

\reference{}Barway, S., Wadedekar, Y., Vaghmare, K., \& Kembhavi, A. K. 2013, MNRAS, 432, 430

\reference{}Battinelli, P., \& Demers, S. 2005, A\&A, 434, 657

\reference{}Binggeli, B., Sandage, A., \& Tammann, G. A. 1985, AJ, 90, 1681

\reference{}Binggeli, B., Tammann, G. A., \& Sandage A. 1987, AJ, 94, 251

\reference{}Black, J. H., \& van Dishoeck, E. F. 1987, ApJ, 322, 412

\reference{}Boker, T., Sarzi, M., McLaughlin, D. E., van der Marel, R. P., Rix, H-W, Ho, L. C., \& Shields, J. C. 2004, AJ, 127, 105

\reference{}Bohringer, H., Briel, U. G., Schwarz, R. A., Voges, W., Hartner, G., \& Trumper, J. 1994, Nature, 368, 828

\reference{}Boselli, A., \& Gavazzi, G. 2006, PASP, 118, 517

\reference{}Boselli, A., Fossati, M., Longobardi, A., et al. 2020, A\&A, 634, L1

\reference{}Bowen, G. H., \& Willson, L. A. 1991, ApJ, 375, L53

\reference{}Butcher, H., Oemler, A. Jr 1984, ApJ, 285, 426

\reference{}Byler, N., Dalcanton, J. J., Conroy, C., \& Johnson, B. D. 2017, ApJ, 840, 44

\reference{}Cappellari, M., McDermid, R. M., Alatalo, K. et al. 2013, MNRAS, 432, 1862

\reference{}Carollo, M., Stiavelli, M., \& Mack, J. 1998, AJ, 116, 68

\reference{}Chabrier, G. 2001, ApJ, 554, 1274

\reference{}Cordier, D., Pietrinferni, A., Cassisi, S., \& Salaris, M. 2007, AJ, 133, 468

\reference{}Choi, J., Conroy, C., Moustakas, J., et al. 2014, ApJ, 792, 95

\reference{}Conroy, C., \& van Dokkum, P. 2012, ApJ, 747, 69

\reference{}Conroy, C., Dutton, A., Graves, G. J., Mendel, J. T., van Dokkum, P. G. 2013, ApJ, 776, L26

\reference{}Conroy, C., Villaume, A., van Dokkum, P. G., \& Lind, K. 2018, ApJ, 854, 139

\reference{}Cordier, D., Pietrinferni, A., Cassisi, S., \& Salaris, M. 2007, AJ, 133, 468

\reference{}C\^{o}t\'{e}, P., Piatek, S., Ferrarese, L., et al. 2006, ApJS, 165, 57

\reference{}Crowl, H. C., \& Kenney, J. D. P. 2008, AJ, 136, 1623

\reference{}Dahmer-Hahn, L. G., Riffel, R., Rodriguez-Ardila, A., et al. 2018, MNRAS, 476, 4459

\reference{}Davidge, T. J., 1990, AJ, 99, 561

\reference{}Davidge, T. J., 1992, AJ, 103, 1512

\reference{}Davidge, T. J., 2014, ApJ, 791, 66

\reference{}Davidge, T. J., 2015a, ApJ, 799, 97

\reference{}Davidge, T. J., 2015b, ApJ, 811, 133

\reference{}Davidge, T. J., 2016, ApJ, 818, 142

\reference{}Davidge, T. J., 2018a, AJ, 156, 233

\reference{}Davidge, T. J., 2018b, RNAAS, 2, A206

\reference{}Davidge, T. J., 2019, AJ, 158, 90

\reference{}Davidge, T. J., \& Courteau, S. 2002, AJ, 123, 1438

\reference{}De Lucia, G., Weinmann, S., Poggianti, B. M., Aragon-Salamanca, A., \& Zaritsky, D. 2012, MNRAS, 423, 1277

\reference{}Denissenkov, P. A., \& Hartwick, F. D. A. 2014, MNRAS, 437, L21

\reference{}Dressler, A. 1980, ApJ, 236, 351

\reference{}Dressler, A., Oemler, A. Jr., Couch, W. J., et al. 1997, ApJ, 490, 377

\reference{}Eikenberry, S., Elston, R., Raines, S. N., et al. 2004, Proc. SPIEE, 5492, 1196

\reference{}Ferreras, I., La Barbera, F., de La Rosa, I. G., Vazdekis, A., de Carvalho, R. R., Falcon-Barroso, J., \& Ricciardelli, E. 2013, MNRAS, 429, L15

\reference{}Ford, H. A., \& Bregman, J. N. 2013, ApJ, 770, 137

\reference{}Fossati, M., Gavazzi, G., Savorgnan, G., et al. 2013, A\&A, 553, A91

\reference{}Fraser-McKelvie, A., Brown, M. J. I., Pimbblet, K., Dolley, T., \& Bonne, N. J. 2018, MNRAS, 474, 1909

\reference{}Gallazzi, A., Charlot, S., Brinchmann, J., White, S. D. M., \& Tremonti, C. A. 2005, MNRAS, 362, 41

\reference{}Geargiev, I. V., \& Boker, T. 2014, MNRAS, 441, 3570

\reference{}Gieles, M., Charbonnel, C., Krause, M. G. H., et al. 2018, MNRAS, 478, 2461

\reference{}Giradi, L., Bressan, A., Bertelli, F., \& Chiosi, C. 2000, A\&AS, 141, 371

\reference{}Gonzalez Delgado, R. M., Perez, E., Cid Fernandes, R., et al. 2017, A\&A, 607, A128

\reference{}Goddard, D., Thomas, D., Maraston, C., et al. 2017, MNRAS, 466, 4731

\reference{}Greenhouse, M. A., Woodward, C. E., Thronson Jr, H. A., Rudy, R. J., Rossano, G. S., Erwin, P., \& Puetter, R. C. 1991, ApJ, 383, 164

\reference{}Groenewegen, M. A. T., \& Sloan, G. C. 2018, A\&A, 609, A114

\reference{}Grootes, M. W., Tuffs, R. J., Popescu, C. C., et al. 2017, AJ, 153, 111

\reference{}Gunn, J. E., \& Gott, J. R. I. 1972, ApJ, 176, 1

\reference{}Hogg, D. W., Blanton, M. R., Brinchmann, J., et al. 2004, ApJ, 601, L29

\reference{}Hou, A., Parker, L. C., \& Harris, W. E. 2014, MNRAS, 442, 406

\reference{}James, P. A., \& Percival, S. M. 2018, MNRAS, 474, 3101

\reference{}Jarrett, T. H., Chester, T., Cutri, R., Schneider, S. E., \& Huchra, J. P. 2003, AJ, 125, 525

\reference{}Karachentsev, I. D., Makarova, L. N., Tully, R. B., Rizzi, L., \& Shaya, E. J. 2018, ApJ, 858, 62

\reference{}Kauffmann, G., White, S. D. M., Heckman, T. M., et al. 2004, MNRAS, 353, 713

\reference{}Khoperskov, S., Haywood, M., Di Matteo, P., Lehnert, M. D., \& Combes, F. 2018, A\&A, 609, A60

\reference{}Kim, S., Rey, S.-C., Bureau, M., et al. 2016, ApJ, 833,207

\reference{}Kim, S., Jeong, H., Lee, J., Lee, Y., Joo, S-J, Kim, H-S, \& Rey, S-C 2018, ApJ, 860, L3

\reference{}Kobayashi, C., Umeda, H., Nomoto, K., Tominaga, N., \& Ohkubo, T. 2006, ApJ 653, 1145

\reference{}Koopmann, R. A., \& Kenney, J. D. P. 1998, ApJ, 497, L75

\reference{}Koopmann, R. A., \& Kenney, J. D. P. 2004a, ApJ, 613, 851

\reference{}Koopmann, R. A., \& Kenney, J. D. P. 2004b, ApJ, 613, 866

\reference{}Koopmann, R. A., Haynes, M. P., \& Catinella, B. 2006, AJ, 131, 716

\reference{}Kwak, S., Kim, W.-T., Rey, S. C., \& Kim, S. 2017, ApJ, 839, 24

\reference{}Kwak, S., Kim, W.-T., Rey, S. C., \& Quinn, T. R. 2019, ApJ, 887, 139

\reference{}Lamperti, I., Koss, M., Trakhtenbrot, B., et al. 2017, MNRAS, 467, 540

\reference{}Larkin, J. E., Armus, L., Knop, R. A., Soifer, B. T., \& Matthews, K. 1998, ApJS, 114, 59

\reference{}Lee, M. G., Kim, M., Sarajedini, A., Geisler, D., \& Gieren, W. 2002, ApJ, 565, 959

\reference{}Li, H., Mao, S., Cappellari, M., et al. 2018, MNRAS, 476, 1765

\reference{}Lisker, T., Grebel, E. K., \& Binggeli, B. 2006a, AJ, 132, 497

\reference{}Lisker, T., Glatt, K., Westera, P., Grebel, E. K. 2006b, AJ, 132, 2432

\reference{}Magrini, L., Coccato, L., Stanghellini, L., Casasola, V., \& Galli, D. 2016, A\&A, 588, A91

\reference{}Maraston, C. 2005, MNRAS, 362, 799

\reference{}Masters, K. L., Mosleh, M., Romer, A. K., et al. 2010, MNRAS, 405, 783

\reference{}McDonald, M., Courteau, S., \& Tully, R. B. 2009, MNRAS, 393, 628

\reference{}McDonald, M., Courteau, S., Tully, R. B., \& Roediger, J. 2011, MNRAS, 414, 2015

\reference{}McGee, S., Balogh, M. L., Bower, R. G., Font, A. S., \& McCarthy, J. G. 2009, MNRAS, 400, 937

\reference{}McQuinn, K. B. W., Skillman, E. D., Cannon, J. M., et al. 2010, ApJ, 724, 49

\reference{}Mei, S., Blakeslee, J. P., C\^{o}t\^{'}, P., et al. 2007, ApJ, 655, 144

\reference{}Meneses-Goytia, S., Peletier, R. F., Trager, S. C., \& Vazdekis, A. 2015, A\&A, 582, A97

\reference{}Meyer, R. E., Sivanandam, S., Moon, D-S. 2019, ApJ, 875, 151

\reference{}Miner, J., Rose, J. A., \& Cecil, G. 2011, ApJ, 727, L15
 
\reference{}Moore, B., Katz, N., Lake, G., Dressler, A., \& Oemler, A. Jr. 1996, Nature, 379, 613

\reference{}Pietrinferni, A, Cassisi, S., Salaris, M., \& Castelli, F. 2004, ApJ, 612, 168

\reference{}Pilkington, K., Few, C. G., Gibson, B. K., et al. 2012, A\&A, 540, A56

\reference{}Postman, M., \& Geller, M. J. 1984, ApJ, 281, 95

\reference{}Puxley, P. J., Hawarden, T. G., \& Mountain, C. M. 1990, ApJ, 364, 77

\reference{}Rayner, J. T., Cushing, M. C., \& Vacca, W. D. 2009, ApJS, 185, 289

\reference{}Riguccini, L., Temi, P., Amblard, A., Fanelli, M., \& Brighenti, F. 2015, ApJ, 810, 138

\reference{}Rock, B., Vazdekis, A., Ricciardelli, E., Peletier, R. F., Knapen, J. H., \& Falcon-Barroso, J. 2016, A\&A, 589, 73

\reference{}Rock, B., Vazdekis, A., La Barbera, F., Peletier, R. F., Knapen, J. H., Allende-Prieto, C., \& Aguado, D. S. 2017, MNRAS, 472, 361

\reference{}Roediger, J. C., Courteau, S., Sanchez-Blazquez, P., \& McDonald, M. 2012, ApJ, 758, 41
 
\reference{}Rosani, G., La Barbera, F., Ferreras, J., \& Vazdekis, A. 2018, MNRAS, 476, 5233

\reference{}Roskar, R., Debattista, V. P., Quinn, T. R., Stinson, G. S., \& Wadsley, J. 2008, ApJ, 684, L79

\reference{}Rudnick, G., Jablonka, P., Moustakas, J., et al. 2017, ApJ, 850, 181

\reference{}Ruggiero, R., \& Lima Neto, G. B. 2917, MNRAS, 468, 4107

\reference{}Saviane, I., Ivanov, V. D., Held, E. V., Alloin, D., Rich, R. M., Bresolin, F., \& Rizzi, L. 2008, A\&A, 487, 901

\reference{}Schawinski, K., Urry, C. M., Simmons, B. D., et al. 2014, MNRAS, 440, 889

\reference{}Schodel, R., Feldmeier, A., Kunneriath, D., Stolovy, S., Neumayer, N., Amaro-Seoane, P., \& Nishiyama, S. 2014, A\&A, 566, A47

\reference{}Schreiber, L., Greggio, L., Falomo, R., Fantinel, D., \& Uslenghi, M. 2014, MNRAS, 437, 2966

\reference{}Seth, A. C., Cappellari, M., Neumayer, N., et al. 2010, ApJ, 714, 713

\reference{}Skillman, E. D., Kennicutt Jr, R. C., Shields, G. A., \& Zaritsky, D. 1996, ApJ, 462, 147

\reference{}Sorce, J. G., Courtois, H. M., Tully, R. B., et al. 2013, ApJ, 765, 94

\reference{}Soto, K. T., \& Martin, C. L. 2010, ApJ, 716, 332

\reference{}Spiniello, C., Trager, S., Koopmans, L. V. E., \& Conroy, C. 2014, MNRAS, 438, 1483

\reference{}Sugai, H., Malkan, M. A., Ward, M. J., Davies, R. I., \& McLean, I. S. 1997, ApJ, 481, 186

\reference{}Taranu, D. S., Hudson, M. J., Balogh, M. L., Smith, R. J., Power, C., Oman, R. A., \& Krane, B. 2014, MNRAS, 440, 1934

\reference{}Temi, P., Brighenti, F., \& Mathews, W. G. 2009, ApJ, 707, 890

\reference{}Trager, S. C., Faber, S. M., Worthey, G., \& Gonzalez, J. J. 2000, AJ, 120, 165

\reference{}Tully, R. B., \& Shaya, E. J. 1984, ApJ, 281, 31

\reference{}Urich, L., Lisker, T., Janz, J., et al. 2017, A\&A, 606, A135

\reference{}van Dokkum, P., Conroy, C., Villaume, A., Brodie, J., \& Romanowsky, A. J. 2017, ApJ, 841, 68

\reference{}Vassiliadis, E., \& Wood, P. R. 1993, ApJ, 413, 641

\reference{}Ventura, P., \& D'Antona, F. 2011, MNRAS, 410, 2760

\reference{}Vera, M., Alonso, S., \& Coldwell, G. 2016, A\&A, 595, A63

\reference{}Wetzel, A. R., Tinker, J. L., Conroy, C., \& van den Bosch, F. C. 2013, MNRAS, 432, 336

\reference{}Willson, L. A. 2000, ARA\&A, 38, 573

\reference{}Zibetti, S., Gallazzi, A., Charlot, S., Pierini, D., \& Pasquali, A. 2013, MNRAS, 428, 1479

\reference{}Zieleniewski, S., Houghton, R. C. W., Thatte, N., Davies, R. L., \& Vaughan, S. P. 2017, MNRAS, 465, 192
 
\end{references}
\end{document}